# Quantifying uncertainty in spatio-temporal changes of upper-ocean heat content estimates: an internationally coordinated comparison

Abhishek Savita[1,2,3], Catia M. Domingues[1,3,4], Tim Boyer [5], Viktor Gouretski[6], Masayoshi Ishii[7], Gregory C. Johnson[8], John M. Lyman[8,9], Josh K. Willis[10], Simon J. Marsland[1,2,3], William Hobbs[1,3,11], John A. Church[12], Didier P. Monselesan[2], Peter Dobrohotoff[1,2], Rebecca Cowley[2], Susan E. Wijffels[2,13]

[1]Institute for Marine and Antarctic Studies (IMAS), University of Tasmania, Tasmania, Australia
[2]Commonwealth Scientific and Industrial Research Organisation, Oceans and Atmosphere, Victoria/Tasmania, Australia
[3]Centre of Excellence for Climate Extremes (CLEX), Australian Research Council, Tasmania, Australia
[4]National Oceanography Centre, Southampton, UK
[5]NOAA/National Centers for Environmental Information, Silver Spring, Maryland
[6]Center for Earth System Research and Sustainability, CliSAP, Integrated Climate Data Center, University of Hamburg, Hamburg, Germany
[7]Meteorological Research Institute, Japan Meteorological Agency, Tsukuba, Japan
[8]NOAA/Pacific Marine Environmental Laboratory, Seattle, Washington
[9]Joint Institute for Marine and Atmospheric Research, University of Hawaiʻi at Mānoa, Honolulu, Hawaii
[10]Jet Propulsion Laboratory, California Institute of Technology, Pasadena, California
[11]Australian Antarctic Program Partnership (AAPP), Tasmania, Australia
[12]Climate Change Research Centre, University of New South Wales, Sydney, Australia
[13]Woods Hole Oceanographic Institution, Woods Hole, U.S.A.

**Corresponding Author**: Abhishek Savita
**Email:** abhishek.abhisheksavita@utas.edu.au

**ORCID numbers:**

Abhishek Savita: 0000-0003-2800-3636
Catia M. Domingues: 0000-0001-5100-4595
Tim Boyer: 0000-0002-8827-4836
Viktor Gouretski:
Masayoshi Ishii: 0000-0002-2564-5548
Gregory C. Johnson: 0000-0002-2985-7732
John M. Lyman:
Josh K. Willis: 0000-0002-4515-8771
Simon J. Marsland: 0000-0002-5664-5276
William Hobbs: 0000-0002-2061-0899
Didier P. Monselesan: 0000-0002-0310-8995
Peter Dobrohotoff: 0000-0001-7315-042X
Susan E. Wijffels: 0000-0002-4717-0811
Rebecca Cowley: 0000-0001-8541-3573
John A. Church: 0000-0002-7037-8194

# ABSTRACT

The Earth system is accumulating energy due to human-induced activities. More than 90% of this energy has been stored in the ocean as heat since 1970, with ~64% of that in the upper 700 m. Differences in upper ocean heat content anomaly (OHCA) estimates, however, exist. Here, we evaluate spread in upper OHCA estimates arising from choices in instrumental bias corrections and mapping methods, in addition to the effect of using a common ocean mask. The same dataset was mapped by six research groups for 1970–2008, with six instrumental bias corrections applied to expendable bathythermograph (XBT) data. We find that use of a common ocean mask may impact estimation of global OHCA by 2–13%. Uncertainty due to mapping method dominates over XBT bias correction at a global scale and is largest in the Indian Ocean and in the eddy-rich and frontal regions of all basins. Uncertainty due to XBT bias correction is largest in the Pacific Ocean within 30°N–30°S. In both mapping and XBT cases, spread is higher since the 1990s. Important differences in spatial trends among mapping methods are found in the well-observed Northwest Atlantic and the poorly-observed Southern Ocean. Although our results cannot identify the best mapping or bias correction schemes, they identify where and when greater uncertainties exist, and so where further refinements may yield the largest improvements. Our results highlight the need for a future international coordination to evaluate performance of existing mapping methods.

# 1. Introduction

The Earth system is accumulating energy due to sustained increase in concentrations of atmospheric greenhouse gases associated with human-induced activities (Le Quéré et al. 2018). Since 1971, more than 90% of this energy has been stored in the ocean as heat, with 64% of that in the upper 700 m (Rhein et al. 2013). Upper-ocean heat uptake delays surface warming (Armour et al. 2013; Otto et al. 2013; Raper et al. 2002) but causes a wide range of adverse impacts (IPCC SROCC: Summary for Policymakers 2019; Stocker 2015), from degradation of marine ecosystems (Bindoff et al. 2019; Hughes et al. 2018; Olsen et al. 2018) to mean sea level rise and associated extreme events (Church et al. 2013; Oppenheimer et al. 2019; Seneviratne et al. 2012; WCRP Global Sea Level Budget 2018).

Multidecadal estimates of OHCA – which provide evidence of physical changes in the ocean, mean sea level and climate – are required along with Coupled Model Intercomparison Project (CMIP) simulations (Eyring et al. 2016) to attribute the detected changes to natural and anthropogenic radiative forcing (Bilbao et al. 2019; Gleckler et al. 2012; Marcos et al. 2017; Slangen et al. 2014; Tokarska et al. 2019) and to constrain uncertainties in CMIP projections used in policy-making and risk management (Carson et al. 2019; SROCC: Summary for Policymakers 2019; van de Wal et al. 2019).

The first observational estimate of multidecadal increase in global OHCA was compiled by Levitus et al. (2000). Since then, a number of regularly gridded OHCA estimates are produced by different groups (e.g. Boyer et al. 2016; Johnson et al. 2019; Meyssignac et al. 2019). Following the findings of Gouretski and Koltermann (2007), these estimates include a diversity of instrumental bias corrections applied to a large portion of the historical ocean temperature profiles, collected by eXpendable Bathythermographs (XBTs) (Abraham et al. 2013). With a significant reduction of systematic depth errors in XBT data (Wijffels et al. 2008), Domingues et al. (2008) showed that the rate of multidecadal increase in global upper OHCA and associated thermal expansion was faster than previously reported in the Intergovernmental Panel for Climate Change (IPCC) Assessment Report 4 (Bindoff et al. 2007). They also showed that CMIP simulations, including both natural (e.g. solar and volcanic) and anthropogenic (e.g. aerosols and greenhouse gases) forcing were in good agreement with their improved observational estimate. Using this improved OHCA (thermal expansion) estimate, Church et al. (2011) were able to satisfactorily account for the processes causing global mean sea level rise since 1971.

All global OHCA (thermal expansion) estimates show a sustained and statistically significant ocean warming (thermosteric sea level rise) in the upper 700 m, since 1971 (or 1993) (Bindoff et al. 2019; Johnson et al. 2019; Oppenheimer et al. 2019; Rhein et al. 2013; WCRP Global Sea Level Budget 2018), despite their differences in instrumental bias correction, baseline climatology and method used to map the uneven spatio-temporal coverage of ocean temperature profiles into regular fields. These differences in estimation procedures, however, introduce uncertainty in linear rates and spatio-temporal evolution of upper OHCA (Abraham et al. 2013; Boyer et al. 2016; Cheng et al. 2016; Good 2017; Lyman et al. 2010; Meyssignac et al. 2019; Palmer et al. 2010).

Initially, quality control and XBT bias corrections were considered the largest source of uncertainty in global OHCA estimates for 1993–2008 in the upper 700 m (Lyman et al. 2010). A later coordinated study (Boyer et al. (2016) demonstrated that, on average, mapping method is the largest source of uncertainty for 1993–2008, and also over the longer 1970–2008 period, followed by spread due to XBT bias correction. This coordinated study additionally reported small differences in global upper OHCA due to varying definitions in ocean mask for two of their eight mapped estimates, based on one of their six XBT bias corrections (their Fig. 2). While differences in spatio-temporal patterns of OHCA change have been noticed, Boyer et al. (2016) did not quantify the relative contribution of influencing factors.

In this study, we extend Boyer et al. (2016) analysis by using the same gridded datasets for the upper 700 m, produced for their coordinated intercomparison, to investigate sensitivity of: (1) global OHCA estimates to the definition of a common ocean mask, and (2) spatio-temporal changes in OHCA to (i) XBT bias correction and (ii) mapping method. We do not investigate spread due to baseline climatology because the required combinations of gridded datasets were not available from all research groups. In addition, we could not include two estimates used in Boyer et al. (2016), as one of them is a global integral and the other was lost due to hard disk failure. Since all of the OHCA estimates mapped the same global Argo dataset from 2005 onwards, and spread is significantly smaller (Boyer et al. 2016), we largely focus on the period 1970 (or 1993) to 2004, when XBTs were the major instrument type (Abraham et al. 2013). **Section 2** provides an overview of the datasets and approaches. **Section 3** presents results from our sensitivity analyses. Discussion and conclusions follow in **Sections 4 and 5**.

## 2. Data description and coordinated approach

*a. Temperature data and intercomparison protocol*

We analyzed the same mapped OHCA dataset versions, depth-integrated for the upper 700 m (**Table 1**), as used in Boyer et al. (2016; their Table 1) with two exceptions: (i) the representative mean from the Pacific Marine Environmental Laboratory (PMEL-R, Lyman and Johnson 2008), which is only a global integral (i.e. no regional fields), and (ii) the Gouretski (2012) estimate (GOU), lost after hard disk failure.

The mapped OHCA dataset versions (**Table 1**) comprise in situ ocean temperature profiles from bottles, Conductivity-Temperature-Depth (CTDs) and XBTs in the EN3v2a global database for 1970–2004 (Ingleby and Huddleston 2007) combined with Argo float profiles for 2000–2008 (Barker et al. 2011). The in-situ profiles were mapped by six research groups, including seven permutations of the XBT data, to account for six bias corrections plus an uncorrected case, totaling 42 dataset members. This dataset combination was specifically produced for the coordinated intercomparison in Boyer et al. (2016), and is not necessarily the same routinely used by the research groups, particularly because global databases are dynamic (i.e. continuously evolving in number of profiles, metadata recovery, quality control, exact and near-duplicates removal, bias corrections, etc). For instance, see the IQuOD Project at www.iquod.org.

As described in Boyer et al. (2016), once the in-situ temperature profiles from the combined global profile datasets were assembled, they were converted into potential temperature anomalies relative to a monthly mean climatology. Anomaly profiles were depth-integrated for two levels, 0–300 m and 300–700 m and then distributed and separately gridded by each research group using their respective mapping methods, and subsequently added together to obtain gridded anomalies for 0–700 m. To convert potential temperature anomalies into OHCA, the gridded data were multiplied by seawater density (1020 kg m$^{-3}$) and heat capacity (4187 J Kg$^{-1}$ $^o$C$^{-1}$) constants.

The 42 OHCA members (**Table 1**) were mapped based on six mapping methods: DOM (Domingues et al. 2008), LEV (Levitus et al. 2012), ISH (Ishii and Kimoto 2009), EN (Ingleby and Huddleston 2007), PMEL (Lyman and Johnson 2008), and WIL (Willis 2004). All OHCA estimates are for 1970–2008, except WIL that starts in 1993, as their mapping method relies on regressions with sea level from satellite altimeter (WCRP Global Sea Level Budget 2018). A summary of the individual mapping methods is found in Boyer et al. (2016).

All gridded OHCA dataset versions in **Table 1** are relative to the same monthly mean baseline climatology from Alory et al. (2007), corresponding to the "C1_H (or H)" case in Boyer et al. (2016). This climatology comprises bottle and CTD profiles from the EN3v1d (the immediate previous version to EN3v2a) database for 1970–2004 (Ingleby and Huddleston 2007) merged with Argo profile floats for 2000–2008 (Barker et al. 2011), and deliberately excludes XBT profiles. Note that, although Boyer et al. (2016) tested the effect of three different climatologies, this was only done for a small subset of eight estimates with a single XBT bias correction (W08; Wijffels et al. 2008) due to limitations in resources. Other insights on the influence of climatology choices are found in Good (2017) and Lyman and Johnson (2014).

The six (out of ten or more) proposed XBT bias corrections used in this study (**Table 1**) may not correct for all recommended temperature and depth factors and may not apply for all types of XBTs manufactured over the years, as explained in Cheng et al. (2016). The large number of proposed XBT bias corrections partly reflects the difficulties imposed by missing metadata in about 50% of XBT profiles (Abraham et al. 2013). To reduce this problem, Palmer et al. (2018) developed a deterministic approach to intelligently assign a set of plausible metadata information to XBT profiles, as part of the IQuOD Project (www.iquod.org). Probabilistic efforts underpinned by machine learning are also being explored in IQuOD (Leahy et al. 2018).

The ten or more XBT bias corrections have been applied to temperature profiles in global databases widely used since 2013 (e.g. WOD13, Boyer et al. 2013; EN4, Good et al. 2013). More recently, Cheng et al. (2018) developed a set of metrics to evaluate the XBT bias corrections and reported that four (Cheng and Zhu 2014 (CH14); Gouretski 2012 (GK12); Gouretski and Reseghetti 2010 (GR10); Levitus et al. 2009 (L09)), out of the ten corrections they examined, appear to be the best performing schemes. Both L09 and GK12 are considered in this study (**Table 1**), along with W08 (Wijffels et al. 2008), I09 (Ishii and Kimoto 2009), GD11 (Good 2011) and CW13 (Cowley et al. 2013). Our results do not fully support the Cheng et al. (2018) conclusions (**Sections 2d and 4**). Note that, their recommendation and associated CH14 bias correction were not available at the time our coordinated OHCA estimates were produced. Only XBT profiles from the recent release of the World Ocean Database (WOD18, Boyer et al. 2018) as well as from IQuOD's first interim release (v01, IQuOD team 2018) use the CH14 correction, as recommended by the XBT science community (Cheng et al. 2018; Goni et al. 2019).

*b. Post-processing and common ocean mask*

Because the 0–700 m OHCA datasets mapped by the six research groups were not necessarily computed with the same temporal and spatial resolutions, we post-process them for the intercomparison in **Section 3.** Our analyses are based on OHCA annual means, interpolated onto a 1° by 1° spherical grid, area-weighted and relative to an ocean mask (65°N-65°S) common to all mapped fields from the six groups (**Fig. 1a**). The original masks from each group are shown in **Fig. 1c-h**, along with bathymetry (**Fig. 1b**), and surface areas listed in the caption. Most differences are in marginal seas (especially the Indonesian Throughflow region) and shelf areas along the west boundary margins of the North Pacific and Atlantic Ocean (particularly off South America), shown as boxes in **Fig. 1**.

Global estimates were calculated by integrating weighted OHCA for all grid points within the common ocean mask, from 1970 (or 1993) to 2008, and its influence is described in **Section 3a.** Basin-integrals follow the color-coded areas in **Fig. 1a.** Our Southern Ocean definition is poleward of 35°S (not shown).

*c. 0-700 m subsampling*

To compare the OHCA estimates mapped by the six research groups with the original unmapped datasets for each XBT bias correction, we subsampled the mapped fields ("*subsampled profiles*") only where temperature profiles were available from surface to 700 m ("*observed profiles*"). Note that, when more than one 0–700 m profile was available within a 1° x 1° grid box for a certain month and year, they were averaged using a median to create one 'superobs' for the location, prior to distribution to the research groups for mapping. As the 0–300 m profiles can be due to a combination of shallow and deep profiles, they were not used for the subsampling exercise. The subsampling was carried out for mapped fields with the common ocean mask applied. In other words, although the mapped estimates contain grid boxes where observed profiles were available: (i) for the upper 300 m (0–300 m); (ii) for the upper 700 m (0–700 m); or (iii) none at all ("*infilled*"), only grid boxes that matched (ii) were selected for the subsampling. The OHCA estimates based on both "*subsampled*" and "observed" profiles were area-weighted prior to integrating them globally (**Section 3d, Figs. 7-9**).

*d. Statistical calculations*

We largely focused on two periods, 1970–2004 and 1993–2004. Spread in OHCA estimates after 2004 is much reduced as it is only due to mapping of Argo data (Boyer et al. 2016; Ishii et al. 2017). The spread due to the XBT bias corrections (**Section 3b**) was calculated on a yearly basis by taking the standard deviation (STD) of the datasets with the six XBT bias corrections (excluding the uncorrected version) for each of the six mapping methods (**Table 1**). We also estimated the average STD for 1970–2004 and 1993–2004, and in some cases, we calculated the STD for the pre-satellite altimeter era (1970–1993). The satellite altimeter era (1993–2004) coincides with an increase in number of deeper XBT profiles (700 m or deeper) during the World Ocean Circulation Experiment (Abraham et al. 2013; Gould et al. 2013; Wijffels et al. 2008). Finally, we averaged the STD due to XBT bias corrections obtained for each of the six mapping methods together to obtain an ensemble mean spread (EnSTD). Spread due to mapping method (**Section 3c**) followed the above, except that STD calculations were computed among mapping methods for each of the six XBT corrections.

Linear trends were calculated using ordinary least squares regression for 1970–2004 and 1993–2004 (**Section 3e**). As we applied a common mask to the OHCA estimates (**Fig. 1a**), unlike Boyer et al. (2016), we expect some differences in terms of global trends. The standard errors (SE) for the linear trends considered autocorrelation and were computed by the variance of the residuals about the fit, following Santer et al. (2008, their equation 4-6).

## 3. Results

### *a. Effect of common ocean mask on global estimates*

The common mask in **Fig. 1a** represents the global ocean domain intersected by the original masks from the six research groups (**Fig. 1c-h;** surface areas listed in the caption). The intersected domain is largely determined by the most conservative mask from ISH (**Fig. 1e**) which excludes the greatest amount of combined ocean area, notably within marginal seas and shelf zones. Consequently, and independent of XBT bias correction, differences between the original and common mask is smallest for ISH (**Fig. 2i**). The largest difference is found for DOM (**Fig. 2g**), about six times larger than the other four original masks (LEV, EN, PMEL and WIL) with similar ocean domain (**Fig. 1**). The ensemble mean differences are about 2% for ISH (**Fig. 2c**) and 13% for DOM (**Fig. 2a**). Potential reasons are discussed in **Section 4**.

Global OHCA differences due to the common mask (individual minus common), vary with XBT bias correction but generally tend to increase with time for all mapping methods, particularly after 1990 (**Fig. 2**, right panels). Compared to the Boyer et al. (2016) results that were based on individual masks rather than a common mask, smaller multidecadal increases in global OHCA are seen in this study for all estimates (**Fig. 2,** left panels), although with similar short-term variability. The effect of the common mask is further reflected in the estimation of the global trends in **Section 3f**.

In the rest of this paper we only use the common ocean mask definition for our global, basin and regional analyses.

*b. Spread due to XBT bias correction*

Global STD maps in **Fig. 3** (left panels) show the spread in gridded OHCA regional patterns in the upper 700 m due to the six choices of XBT bias corrections for timeseries starting in 1970 (DOM, LEV, ISH, EN, PMEL; **Table 1**). EnSTD maps were estimated by averaging the global STD patterns across the five mapping methods (**Fig. 3**, right panels) for three periods: 1970–2004 (longest), 1970–1992 (pre-altimeter era) and 1993–2004 (altimeter era). Inclusion of WIL's estimates in the EnSTD for 1993-2004 (**Fig. 3i**) do not alter the results seen in **Fig. 3h**.

EnSTD is maximum across all ocean basins within 30°N–30°S, with highest values in the Pacific Ocean and for 1993–2004 compared to 1970–1992 (**Fig. 3**, right panels). While the Pacific maximum for 1970–1992 is centralized (**Fig. 3g**), the pattern for 1993–2004 is broken into two zonally-extended cells, found further from the equator (**Fig. 3h**). Consequently, the longest period, 1970–2004, reflects a combination of these EnSTD patterns (**Fig. 3f**).

Over 1970–2004, the maximum in the EnSTD pattern across 30°N–30°S (**Fig. 3f**) is mainly influenced by LEV, ISH and EN (**Fig. 3b,c,d**) and their radii of influence (shape and size). The imprint of the radius of influence used by each of these mapping methods becomes obvious after comparing with PMEL (**Fig. 3e**). PMEL has a clearer delineation of the repeated XBT lines in their STD maps because their choice of physically-based correlation length scales and signal-to-noise ratios (SNR), from their objective mapping, relaxes toward the initial guess of zero anomalies in data-sparse regions (Boyer et al. 2016; Lyman and Johnson 2008).

In contrast, maxima in the STD along XBT lines are not evident in DOM (**Fig. 3**a) but instead found across the Southern Ocean, where XBT measurements are limited to a small number of meridional repeat lines (Goni et al. 2019), and where seawater is colder and vertical temperature stratification is weaker relative to lower latitudes. While all other mapping methods depend on local fitting, DOM minimizes errors at global and local scales simultaneously. For that, DOM relies on statistics from a reduced set of Empirical Orthogonal Functions (EOFs) from satellite altimeter to infill the sparser in-situ ocean temperature observations (e.g. Church et al. 2004; Pittman 2016). The relative discrepancies in STD patterns for DOM are not unexpected as their method minimizes spread due to XBT bias correction locally while also having a far-reaching influence, not necessarily associated with the ocean regions sampled by the XBT profiles.

Overall, per square meter, DOM has the highest sensitivity to the differences in XBT bias corrections, independent of ocean basin, and well above the EnSTD for both 1970–2004 and 1993–2004 (**Fig. 4**, right panels). PMEL is the least sensitive for 1970–2004 but not necessarily for 1993–2004. From a basin perspective, the Pacific Ocean has the highest spread per square meter, followed by the Atlantic and Indian Ocean. DOM is an exception, in which the Pacific and Indian basins have similar sensitivities, around 60 MJ $m^{-2}$ or 6 x $10^{-14}$ ZJ $m^{-2}$; (1 MJ = $10^6$ J; 1 ZJ = $10^{21}$ J), and higher than the Atlantic Ocean. The STD timeseries for individual basins (**Fig. 4**, left panels) are very similar to the global variations reported in Boyer et al. (2016). XBT spread per square meter is higher during 1989–2000 compared to previous years, with a maximum peak around 2000 that decays to zero in 2005, when only Argo data were included in the coordinated protocol (**Section 2a**).

*c. Spread due to mapping method*

Global maps of STD patterns due to the six choices in mapping methods (**Table 1**) are similar across XBT bias corrections (not shown), and so we only present their EnSTD patterns (**Fig. 5**, left panels), obtained by averaging the STD patterns across the six XBT bias corrections. In general, STD maxima largely coincide with well-known highly energetic eddy regions and frontal systems seen in altimeter sea level (Fu et al. 2010), including Gulf Stream and Kuroshio-Oyashio boundary current extensions, Brazil-Malvinas Confluence, Agulhas and East Australian Current retroflections, and along the Antarctic Circumpolar Current, particularly in the Indian sector. In contrast with the EnSTD due to XBT bias correction (Figs. 3h and 3i), the EnSTD due to mapping method increases after inclusion of WIL's estimates in all basins for 1993–2004 (**Fig. 5d** compared to **5c**), also evident in the zonal integrals (**Fig. 5**, right panels).

The zonally-integrated EnSTD have similar patterns across the latitudinal bands for the three time-average periods and is maximum in the Southern Ocean (35°S–60°S) (**Fig. 5**, right panels). The highest STD contributions for the Southern Ocean peak are from the Indian and Atlantic sectors (40°S–50°S), followed by the Pacific sector (50°S–60°S). In the Pacific, the largest peak lies around 0° to 20°N, followed by three secondary peaks, 30°N–40°N, 20°S–40°S and 50°S–60°S. The Atlantic has only one secondary peak (30°N–40°N). The Indian Ocean has a plateau from 10°S to 30°S, with STD values decaying north of 10°S.

Overall, per square meter, the largest spread in OHCA due to mapping method is in the Indian Ocean for all XBT bias corrections, over 1970–2004 and 1993–2004 (**Fig. 6**, right panels). In fact, the Indian Ocean has the highest spread during most years except in the mid-1980s, when the Pacific has two maxima (**Fig. 6**, left panels). Over 1993–2004, the EnSTD for the Indian Ocean is almost twice as large as for the other basins, mainly due to the two maxima in 1997–1998 and 2001–2002, seen across all XBT bias corrections. The EnSTD for the Atlantic Ocean has the smallest mapping spread. In terms of individual XBT bias corrections, L09 and W08 have the highest and lowest spread respectively. Over 1970–2004, the EnSTD difference between the Indian Ocean and the other basins is not as large as during 1993–2004, and the lowest spread is for the global ocean. Note that, as the EnSTD of each basin was calculated individually, their sum is not necessarily equal to the EnSTD for the global integral. Rapid decrease in spread after 2004 is largely associated with increase in Argo floats, with near-global coverage in November 2007 (Riser et al. 2016; Roemmich et al. 2019).

### d. *0–700 m subsampling: effect of mapping methods*

To compare OHCA spread with and without the influence of mapping method, we subsampled the mapped fields ("*subsampled profiles*") for the locations where data were collected from the surface to 700 m ("*observed profiles*"), as explained in **Section 2c**. This subsampling allows us to review how the gridded estimates were modified by the mapping methods (e.g. length scales, smoothing, etc) relative to the original observed profiles. Both "*subsampled*" and "*observed*" data were area-weighted before computing the "*global*" integrals (**Figs. 7-9**). In this case the "*global*" integral only represents a fraction of the observed profiles used in Fig. 2 (i.e. only data that extends from 0–700 m as described in case (ii), **Section 2c**).

The *"observed"* OHCA timeseries and the six corresponding timeseries *"subsampled"* from the mapped fields are shown in **Fig. 7**, together with their differences (*"subsampled"* minus *"observed"*). In **Fig. 7a** the *"observed"* timeseries with the CW13 correction has a large OHCA spike during 1999–2000, even greater than the uncorrected timeseries (No_corr). This spike is largely reduced by the mapping methods in the *"subsampled"* timeseries (**Fig. 7b-g**). In general, the disparity across the XBT corrections is largest in the *"observed"* timeseries during 1990–2004 (**Fig. 7a**), with the same valid for the *"subsampled"* timeseries (**Fig. 7b-g**), as indicated by the differences (**Figs. 7h-m**). The CW13 correction has the largest negative difference, associated with the 1999–2000 spike in the *"observed"* timeseries (**Fig. 7a**). CW13 differs from the other XBT bias correction schemes in that they were based on a small fraction of the global dataset but with the highest quality-controlled XBT/CTD profiles. The under-correction is probably due to a combination of a small number of comparison data in the CW13 study, lack of manual quality-control in the EN3v2a dataset used for this study, missing metadata for XBT types, and the impact of the vertical temperature gradient (dependent on latitude) on depth (fall-rate) and thermal biases.

Agreement among *"subsampled"* and *"observed"* timeseries depends on XBT correction and time-average considered (**Fig. 8**). Generally, LEV and DOM have the largest Root Mean Square Error (RMSE) and minimum correlation with the *"observed"* timeseries for most XBT corrections, PMEL has the smallest RMSE and largest correlation, and the other mapping methods lie in-between. Although L09 and GK12 made the top four recommended XBT bias corrections (Cheng et al. 2018), our Taylor diagrams show these corrections differ from each other for 0–700 m profiles (**Fig. 8c,e**), and that the comparison is sensitive to time period and mapping method used.

*e. 0–700 m subsampling: spread due to XBT bias correction and mapping method*

For the "globally-integrated *subsampled*" timeseries formed by subsampling the mapped fields at the locations of observed 0–700 m profiles (**Section 2c**), OHCA spread due to XBT bias correction (**Fig. 9a**) can be higher than spread due to mapping method (**Fig. 9b**), particularly in the 1990s and in contrast to the full-grid timeseries (**Figs. 4 and 6**). Annual variations in spread due to both mapping method and XBT bias correction in **Fig. 9** are highly correlated with the annual number of 0–700 m profiles prior to 2005 – i.e. higher spread with higher number of profiles or fraction of the global ocean sampled (Boyer et al. 2016, their **Fig. 3**). An exception is when the number of profiles decreases around 1998–2001 while the XBT-related spread increases (**Fig. 9a**). This STD peak reaches 12 ZJ for the timeseries based on the 0-700 m "*observed*" profiles (i.e. not mapped, **Section 2c**) and coincides with a peak in OHCA for the CW13 correction (**Fig. 7a**). After CW13 is excluded from the 0–700 m "*observed*" *profiles* estimation, the STD is halved to 6 ZJ (**Fig. 9a**). From 2005 onwards, XBT-related spread is zero, as there are only Argo floats. On average, mapping-related spread during the Argo period is 2 ZJ compared to 4–5 ZJ during the 1990s (**Fig. 9b**).

Annual maps of 0–700 m *"observed"* profiles (**Fig. 10**) reveal the higher XBT-related spread in "*global*" OHCA during the 1990s (**Fig. 9a**) comes from XBT lines crossing the three basins within 30°S–30°N, and that the maxima during 1998–2001 mainly come from the Pacific. When CW13 is removed from the STD calculations, the Pacific maxima (orange to red) are reduced to the same STD levels (cyan) observed across other years in the 1990s (not shown).

As expected mapping-related spread decreases with improved spatio-temporal coverage during the Argo era (e.g. Johnson et al. 2019, their Fig. 3.6; WCRP 2018, their Fig. 4). Increase in OHCA spread with increase in the number of profiles, however, is counter-intuitive prior to the Argo era (**Fig. 9b**). This direct and counter-intuitive correlation can be explained through annual global maps (**Fig. 11**), where OHCA spread is based only on "*subsampled*" 0–700 m profiles with the L09 correction. Mapping-related spread is largely independent of XBT bias correction at gridscale (not shown), similar to the findings in **Section 3c**. Annual maps reveal the STD maxima originate in eddy-rich regions and frontal systems in both hemispheres, consistent with the patterns of the full-grid in **Fig. 5**. Prior to the 1990s, there is less sampling of the energetic locations causing higher OHCA spread mainly in the northern hemisphere (Gulf Stream and Kuroshio). After the 1990s, there is higher coverage of the energetic systems in the southern hemisphere (Agulhas and East Australian Current retroflection and Brazil-Malvinas Confluence) as well as across the western and equatorial Pacific, another source region of higher spread.

f.  Linear trends

Global OHCA trend maps were estimated with the L09 correction for 1970–2004 (DOM, LEV, ISH, EN, PMEL) and 1993–2004 (plus WIL), including ensemble mean and spread (**Fig. 12**). As reported for OHCA spread due to mapping method (e.g. **Section 3c**), trend maps are insensitive to XBT bias correction at gridscale (not shown).

For 1970–2004, the OHCA trend patterns (and their ensemble mean) are generally consistent across mapping methods, with warming rates everywhere, except in some patches of the tropical-subtropical western Pacific Ocean, southeast Indian Ocean and southwestern Atlantic (**Fig. 12**, left column). Enhanced warming in the North Atlantic and North Pacific western boundary extensions are evident in all mapping methods (and their ensemble mean) except for DOM, which shows a weaker trend, particularly in the North Atlantic. ISH shows a weaker warming trend along the ACC pathway in the Indian sector of the Southern Ocean compared to the other mappings. In the South Pacific, DOM, LEV, EN have some cooling patches that are not evident in ISH. Some differences are seen in the Indian Ocean and South Atlantic for the cooling patches in LEV, ISH and EN (not fully sampled by PMEL) whereas DOM displays basin-scale warming. While DOM and ISH tend to have smoother and more zonally elongated large-scale features, in LEV (followed by EN) the large-scale features are immersed in smaller noise-like features, associated with the shape and size of the radii of influence used in their mapping approaches.

For 1993–2004, the global OHCA trend patterns (and their ensemble mean) are also generally consistent across mapping methods (**Fig. 12**, middle column), and largely resemble altimeter sea level ENSO-like low frequency variability (Hamlington et al. 2020; Lyu et al. 2017). The largest discrepancies are seen in the Southern Ocean and North Atlantic between DOM and the other methods. DOM has the strongest wavelike cooling pattern (interposed by warming south of Australia) and the weakest warming rate in the subpolar North Atlantic. Despite thermosteric variability being a dominant driver of regional sea level (Hamlington et al. 2020; Vivier et al. 2005), DOM cannot rule out the influence of other contributions (Fasullo and Gent 2017; Durack et al. 2014) – such as halosteric contributions to density-compensation (e.g. North Atlantic) and ocean bottom pressure (e.g. Southern Ocean) leading to mass changes – since they use EOFs from altimeter to interpolate across the sparsely in-situ observations.

As expected, maxima in the EnSTD trend patterns are found in the highly energetic eddy and frontal regions in both trend periods (**Fig. 12**, right column) but more obvious over 1993-2004 (see also **Fig. 5**). The highest STD regions adjacent to Antarctica in the southern Indian and Pacific Oceans are largely associated with the trend differences in DOM, as explained above.

While the impact of the various XBT bias corrections is negligible at gridscale, it becomes relevant to linear trends when the gridscale differences are integrated over large domains, from zonal, basin to global, for 1970–2004 (**Fig. 13**) and 1993–2004 (**Fig. 14**). Because of our use of the common ocean mask (**Section 3a**) and slightly shorter periods (up to 2004 rather than 2008), the global OHCA trends here are not expected to be identical to Boyer et al. (2016). For instance, the smallest global trend for 1970–2004 is from PMEL rather than ISH as in Boyer et al (2016).

For 1970–2004 (**Fig. 13**), the two most striking features are from DOM for the various XBT bias corrections: (i) the largest warming trends for the globe, North Pacific and Southern Hemisphere (all basins), in many instances, statistically different from the respective trends from the other mapping methods; and (ii) the smallest warming trends for the North Atlantic, statistically different from the other methods. In general, the smallest OHCA trends are from PMEL (excluding I09 and CW13 for which LEV trends are slightly smaller) because they revert to zero anomalies in data sparse regions, especially in the Southern Hemisphere (**Fig. 12**). The fastest warming rate is for the North Atlantic in all mapping methods (note the different vertical axis scale), where EN has the largest value and DOM the smallest (**Fig. 13**). The North Atlantic is particularly well-observed, so the trend differences might be somewhat surprising, however, spread is large in eddy energetic regions (**Fig. 5**). Except for DOM, warming trends in the poorly-observed Southern Ocean for the various mapping methods (and XBT corrections) are not statistically different from each other.

For 1993–2004 (**Fig. 14**), WIL tends to have the highest warming trends but, in many instances, they are not statistically different from the other methods. The wavelike cooling pattern in **Fig. 12** from DOM explains their cooling trend in the Southern Ocean, which can be statistically different from the other methods depending on XBT correction. The North Indian has a cooling trend for all mapping methods and for most XBT corrections. Overall, the higher range in OHCA trends for 1993–2004 (11 years) compared to 1970–2004 (34 years) can be partially explained by a greater influence of interannual to decadal variability over a shorter period (Johnson and Lyman 2020).

4. **Discussion**

We used the same internationally-coordinated protocol as in Boyer et al. (2016) to quantify the uncertainty (spread) introduced by choices due to six XBT bias corrections (**Section 3b**) and due to six mapping methods (**Section 3c**) for global, basin and regional estimations of OHCA and trends (**Section 3f**), relative to a historical climatology and a common ocean mask (**Section 3a**). Our analysis considered an ensemble with 42 OHCA dataset members, depth-integrated in the upper 700 m, and mapped from 1970 (or 1993) to 2008 (**Table 1**). To compare the influence of the various mapping methods on the observed data (**Section 3d**), we subsampled the mapped estimates where profiles were measured from surface to 700 m (i.e. excluding 0–300 m profiles, **Section 2a**). We also used these subsampled estimates (i.e. with mapping influence) to evaluate spread due to mapping method at the observed locations (**Section 3e**), and together with the observed profile estimates (i.e. without mapping influence) to evaluate spread due to XBT bias correction at sampled locations.

Unlike Boyer et al. (2016), we did not evaluate spread due to choices in climatology, as our gridded OHCA estimates were only relative to one historical climatology due to time and storage constraints. The other four differences from Boyer et al. are: no estimates from Gouretski et al. (2012), as those were lost after hard disk failure; no estimates from PMEL–R, the second mapping step from Lyman and Johnson (2008) because it does not involve regional fields; use of a common ocean mask (**Section 2b**); and estimation periods ending in 2004 (rather than 2008) to coincide with the last year that XBT data were included in the combined global database (**Section 2a**).

The use of a common ocean mask reveals that inclusion/exclusion of shallow seas (**Fig. 1**) may influence global OHCA estimates by up ±13%, equivalent to the contribution from the ocean below 2000 m (Meyssignac et al. 2019), and larger than the ~2% implied for shallow areas in previous studies (Boyer et al. 2016; Meyssignac et al. 2019). Differences depend on mapping method (and XBT bias correction) and tend to increase during the 1990s (**Fig. 2**). The largest difference is for DOM (Church et al. 2004; Domingues et al. 2008), which relies on EOF patterns from satellite altimeter to guide infilling where the coordinated protocol did not include in-situ data across shallow seas (depth <300 or 700 m for 1970–2004 or depth <1500 m after 2004). As a result, DOM can be influenced by the global and regional fittings for 0–300 m and 300–700 m in addition to non-thermosteric signals from shallow areas (Landerer et al. 2007; Wu et al. 2019). These two factors might partially explain the large difference (probably an overestimate) compared to the other groups with similar ocean domains originally (surface areas listed in **Fig. 1** caption). The mapping spread is greatest along coastal boundaries, and the higher spread since the 1990s is likely related to the observed surface ocean warming of WBCs, arising from intensification and poleward shift with a warming climate (Wu et al. 2012; Yang et al. 2016).

We compared OHCA estimates using two of the four XBT bias corrections (L09 and GK12) recommended by Cheng et al. (2018), along with four widely used corrections (W08, I09, GD11, CW13), with and without mapping influence (**Figs. 2, 6-8, 12-14**). The OHCA estimates based on L09 and I09 are very similar to each other at the locations sampled by the 0–700 m profiles (**Figs. 7a, 8b,c**), and the same is valid for GK12 and GD11 (**Figs. 7a, 8d,e**). These pairings also hold for the OHCA values subsampled from the mapped estimates, but to a more variable degree depending on mapping (**Figs. 7, 8**), and to a lesser extent if all gridded values are considered (**Fig. 2**). L09 and GK12 are in the top four recommended by Cheng et al. (2018) but not I09 or GD11. At the same time, we find that L09 and GK12 OHCA estimates are more dissimilar from each other than L09-I09 or GK12-GD11, particularly during the 1990s. In fact, I09 and GD11 were not recommended by Cheng et al. partly for not considering all known factors influencing XBT biases. Despite Cheng et al. (2018) not recommending I09 and GD11, as those corrections did not consider all known factors influencing XBT biases, we find they provide similar results to the corrections that were highly recommended.

To the best of our knowledge, the six XBT bias corrections were correctly implemented in the EN3v2a database (**Section 2a**) but we cannot rule out discrepancies in case of database-dependency (e.g. W08 comparison for EN3v1d and EN3v2a in Boyer et al. 2016; EN4 and WOD13 comparisons in Cheng et al. 2018). Our identified similar pairings (L09-I09 and GK12-GD11) are also evident in Cheng et al. (2016, their Figure 3), who noted that CH14-GK12 are closer to each other than CH14-L09. So, it is possible that correction for all known bias factors, although desirable, may not be the single best indicator for performance and/or that the Cheng et al. (2018) metrics may still be imperfect, having in mind the disagreements they reported for global and side-by-side corrections. The community has made tremendous progress to better understand and develop XBT bias corrections that have greatly reduced artifacts in OHCA estimates but it is likely that uncertainty due to XBT bias correction for global (and basin) integrals may not yet be smaller than reported in Boyer et al. (2016). In sum, our results suggest that it is still important to quantify spread in OHCA estimates (**Figs. 3, 4, 9a, 10-13**) due to more than the top four XBT bias corrections recommended by Cheng et al. (2018), and to have metrics separated by error type and not only XBT type for evaluation of the correction schemes.

Regionally, our results show that the highest spread in OHCA due to XBT bias corrections is within 30°N–30°S, especially in the Pacific and over the 1990s (**Fig. 3**), except for PMEL (physically based smaller radius of influence) and DOM (spread is not necessarily related to XBT data). Spread is smaller in cooler and less stratified waters at high latitudes (e.g. Fig. 9; see also Hutchinson et al. 2013; Ribeiro et al. 2018), although the Southern Ocean is quite undersampled (Abraham et al. 2013). Increase in spread since the 1990s is highly correlated with the number of observed 0–700 m profiles (**Fig. 10**) prior to Argo, due to more widely and deeper sampling of energetic regions in the Southern Hemisphere compared to before. DOM is by far the most sensitive mapping method to spread in XBT corrections (**Fig. 4**).

The large spread we found in eddy-rich regions (**Fig. 5**) is mainly due to mapping method, including the well-observed northwest Atlantic and the poorly-observed Southern Ocean. While WIL can better account for the energetic mesoscale features than LEV, ISH, EN and PMEL (e.g. **Fig. 12**), as resolved by altimeter sea level, this same altimeter variability ("high frequency weather noise") is used in DOM as one of their sources of uncertainty to reconstruct large scale patterns ("low frequency climate signal"). Mesoscale eddy variability (Fu et al. 2010; Hughes and Williams 2010; Penduff et al. 2011) can also affect the slower-varying larger scale climate signals through an inverse cascade of kinetic energy (Penduff et al. 2018, 2019). Our results agree with (Wang et al. 2018), who showed significant spread in the WBC and ACC frontal regions, up to 10 times larger than in other ocean areas. We have shown that the spread due to mapping method is largely independent of XBT bias correction at gridscale resolution. We find that the Indian Ocean has by far the largest basin-scale spread during the 1990s due to mapping method and that spread at observed 0–700 m locations became higher when more extensive deeper sampling took place in the 1990s (**Figs. 3-6, 9b**), with implications for understanding the role of the subsurface ocean in the climate system.

Although Wang et al. (2018) previously reported that the major uncertainty in OHCA basin estimates comes from the Pacific and Southern Oceans, we have distinguished the relative contributions of spread due to XBT bias correction and mapping method per metre square (i.e. independent of basin area). We have shown that the two types of spread appear similarly important for both the Pacific and Atlantic Oceans (e.g. compare **Figs. 4 and 6**), however, XBT spread is largest for the Pacific Ocean while mapping spread is largest for the Indian Ocean, including their respective Southern Ocean sectors.

## 5. Conclusions

Intercomparison of spatio-temporal changes of upper OHCA estimates is generally complicated because studies do not necessarily use the same bias-corrected datasets, baseline climatologies, mapping methods and time periods. With the objective to quantify uncertainty in OHCA due to mapping methods and XBT bias corrections, in addition to the effect of a common mask, we followed a coordinated approach as in Boyer et al. (2016), in which six research groups applied their mapping methods to temperature profiles from a combined global database, for 1970 to 2008, including permutations of the XBT data to account for six (out of ten or more) bias corrections proposed (Cheng et al. 2018). Although our results cannot identify the best mapping or bias correction schemes, they identify where and when greater uncertainties exist, and so where further understanding and refinements may yield the largest improvements.

We recommend that:

- Refinement of XBT bias corrections should focus on profiles within 30°N–30°S, during the 1990s, and particularly from the Pacific Ocean. Correction of XBT biases have not been fully solved yet for OHCA estimation purposes.
- Further refinement of XBT bias corrections should consider the same dataset version for both calibration and benchmarking of schemes. Soon, it will be possible

- to benefit from the first internationally-coordinated quality-controlled ocean database (IQuOD), with assigned uncertainty and intelligent metadata.

- Users should be aware of larger uncertainty, particularly after 1990 and before 2006, in the Pacific Ocean (due to spread in both XBT bias correction and mapping method), in the Indian Ocean (largely due spread in mapping method), and in highly energetic eddy and frontal regions (e.g. where altimeter sea level has the largest variances).

- Future coordinated intercomparisons are necessary to evaluate performance of existing mapping methods (Barth et al. 2014; Cheng et al. 2017; Kuusela and Stein 2018), using synthetic profiles from observations (Argo and/or altimeter) as well as from de-drifted model simulations which conserve tracer properties (Allison et al. 2019; Garry et al. 2019; Good 2017; Palmer et al. 2019).

- Aliasing by mapping methods owing to the historical spatio-temporal undersampling is a limitation that should be recognized, when considering times and ocean volumes over which to estimate variability and trends. Continued recovery of actual profile data and metadata is valuable to reduce uncertainty over the past decades (e.g. Global Oceanographic Data Archeology and Rescue (GODAR), https://www.iode.org). Now and into the future, a sustained observing system fit for monitoring ocean climate change is the critical course of action (e.g. Global Climate Observing System (WMO 2018).

- Lessons can be learnt from the sea surface temperature community who has been subjected to similar problems (e.g. Kennedy 2014) and synergistic efforts should be encouraged (e.g. Atkinson et al. 2014).

- Overall, improved estimation of OHCA and proper quantification of uncertainty are relevant to hindcast reanalysis which assimilate ocean observations (Storto et al.

2019), for ocean and/or climate model analyses to investigate mechanisms of ocean heat uptake (Dias et al. 2020a & b (submitted); Gregory et al. 2016; Couldrey et al. 2020 (submitted)) and to attribute changes to natural and anthropogenic drivers (Gleckler et al. 2012), and ultimately to increase the confidence in projections of climate and sea level change relevant for a large community of policy- and decision-makers.

## 6. Acknowledgements

AS is supported by a Tasmanian Graduate Research Scholarship, a CSIRO-UTAS Quantitative Marine Science top-up and by the Australian Research Council (ARC) (CE170100023; DP160103130). CMD was partially supported by ARC (FT130101532) and the Natural Environmental Research Council (NE/P019293/1). RC was supported through funding from the Earth Systems and Climate Change Hub of the Australian Government's National Environmental Science Program.  TB is supported by the Climate Observation and Monitoring Program, National Oceanic and Atmosphere Administration U.S. Department of commerce. SG is supported by the Joint UK DECC/Defra Met Office Hadley Centre Climate Programme (GA01101). GCJ and JML are supported by NOAA Research and the NOAA Ocean Climate Observation Program. PMEL Contribution Number 5065.  JAC is supported by the Centre for Southern Hemisphere Oceans Research (CSHOR), jointly funded by the Qingdao National Laboratory for Marine Science and Technology (QNLM, China) and the Commonwealth Scientific and Industrial Research Organization (CSIRO, Australia) and Australian Research Council's Discovery Project funding scheme (project DP190101173). Data used in this study are available on request.

# List of tables & figures

## List of Tables

**Table 1.** Sensitivity experiments used to create the 42 OHCA dataset versions accounting for variations in XBT bias corrections (six corrections plus an uncorrected version) and six mapping methods. All experiments are relative to the monthly mean baseline climatology from Alory et al. (2007) and which does not include XBT data. Mapping methods are from: DOM (Domingues et al. 2008), LEV (Levitus et al. 2012), ISH (Ishii and Kimoto 2009), EN (Ingleby and Huddleston 2007), PMEL (Lyman and Johnson 2008) and WIL (Willis et al. 2004). Periods available are for 1970-2008 or 1993-2008, depending on the mapping method.

## List of Figures

**Fig. 1.** Ocean mask definitions and bathymetry. (a) Common mask for global and basins (color). (b) Bathymetry (km) from ETOPO5 (https://doi.org/10.7289/V5D798BF). (c)–(h) Original masks from DOM, LEV, ISH, EN, PMEL and WIL respectively, where blue denotes where there is data coverage and white where there is no data. Southern Ocean basin is poleward of 35°S (not shown). Boxes (dotted line) illustrate major differences in the original masks among research groups. The total area of common mask (2.86 x $10^{14}$ $m^2$) and individual masks are: DOM (3.22 x$10^{14}$ $m^2$), LEV (3.04 x$10^{14}$ $m^2$), ISH (2.91 x $10^{14}$ $m^2$), EN (3.11 x $10^{14}$ $m^2$), PMEL (3.18 x $10^{14}$ $m^2$) and WIL (3.11 x $10^{14}$ $m^2$).

**Fig. 2.** Global OHCA timeseries for six XBT bias corrections as well as uncorrected version (see legend). Left: Timeseries based on the common mask for each mapping method: (a) DOM, (b) LEV, (c) ISH, (d) EN and (e) PMEL for 1970–2008, and (f) WIL for 1993–2008. Right: Differences based on the original mask minus common mask.

Units: ZJ.

**Fig. 3**. OHCA spread due to XBT bias correction. Left: For each mapping method ((a) DOM, (b) LEV, (c) ISH, (d) EN and (e) PMEL) averaged over 1970–2004. Right: Ensemble mean spread (EnSTD) across mapping methods for different time periods (f) 1970–2004, (g) 1970–1992, (h) 1993–2004 excluding WIL (only available since 1993) and (i) 1993–2004 including WIL. Units: ZJ.

**Fig. 4.** Global and basin OHCA spread due to XBT bias correction per metre square. Left: Annual timeseries from 1970–2008 for each mapping method: (a) DOM, (b) LEV, (c) ISH, (d) EN, (e) PMEL and (f) WIL. Middle: Spread averaged over 1970–2004 organised by (g) mapping method, (h) by basin and (i) ensemble mean spread (EnSTD, gray bars) across mapping methods organized by basin. Right: same as middle but for 1993–2004. Units: ZJ m$^{-2}$.

**Fig. 5.** OHCA ensemble mean spread (EnSTD) due to mapping method across XBT bias corrections for different periods. Left: Global patterns for (a) 1970–2004, (b) 1970–1992, (c) 1993–2004 excluding WIL and (d) 1993–2004 including WIL. Right: Zonal integrals for (e) global, (f) Pacific, (g) Atlantic and (h) Indian oceans. Units: ZJ.

**Fig. 6.** Global and basin OHCA spread due to mapping method per metre square. Left: Annual timeseries from 1970–2008 for each XBT bias correction: W08 (a), I09 (b), L09 (c), GD11 (d), GK12 (e) and CW13 (f). Middle: Spread averaged over 1970–2004 organised by XBT correction (g), by basin (h) and the ensemble mean spread (EnSTD,

gray bars) across XBT bias corrections organized by basin (i). Right: Same as middle but for 1993–2004. Units: ZJ m$^{-2}$.

**Fig. 7.** Global OHCA annual timeseries for six XBT bias corrections as well as uncorrected version (see legend) based on subsampled profiles for 0–700 m only. Left: For observed profiles with no mapping applied ("*observed profiles*") and subsampled profiles ("*subsampled*") from each mapping method: (b) DOM, (c) LEV, (d) ISH, (e) EN and (f) PMEL for 1970–2008, and (g) WIL for 1993–2008. Right: Differences based on the original mask minus common mask. Units: ZJ.

**Fig. 8.** Taylor diagrams for each XBT bias correction comparing "*observed profiles*" for 0–700 m (only) and respective subsampled profiles mapped by six methods (color legend). (A): For 1970–2004. (B): For 1993–2004. XBT bias corrections are: (a) W08, (b) I09, (c) L09, (d) GD11, (e) GK12 and (f) CW13. STD (black axis) and Root Mean Square Error (RMSE, green axis) in ZJ. Correlation coefficients (blue axis) are normalized.

**Fig. 9.** Global OHCA spread timeseries based on subsampled profiles for 0–700 m only (left axis), and profile numbers per year (black line with circle, right axis). (a): Spread due to XBT bias correction across six mapping methods ("*subsampled*") and for observed profiles without mapping ("*observed*"), and with and without the XBT correction from CW13. (b) Spread due to mapping methods across six XBT bias corrections. Horizontal lines represent the time mean for 1970–2004 (or 1993–2004 for WIL). Units: ZJ.

**Fig. 10.** Annual global maps of OHCA spread due to XBT bias correction from observed profiles ("*observed*") for 0–700 m only, from 1970 to 2004. Units: ZJ.

**Fig. 11.** Annual global maps of OHCA spread due to mapping method using the L09 XBT bias correction based on subsampled profiles ("*subsampled*") for 0–700 m only, from 1970 to 2004. Units: ZJ.

**Fig. 12.** OHCA linear trend maps based on six mapping methods (DOM, LEV, ISH, PMEL and WIL) using the L09 XBT bias correction and their ensemble mean of the mapping methods. Left: for 1970–2004. Middle: for 1993–2004. Right: Ensemble mean spread (EnSTD) for 1970–2004 (top) and for 1993–2004 (bottom). Units: GJ/$m^2$; 1GJ=$10^9$J.

**Fig. 13.** Linear OHCA trends for different basins (panels) and for each XBT bias correction (x-axis) and for each mapping method (color legend) for 1970–2004. Errorbars take into account the reduction in the degrees of freedom due to the temporal correlation in the residuals (**Section 2d**). Unit: JZ/year.

**Fig. 14.** As in Fig. 13 but for 1993–2004, and including WIL (only available since 1993). Unit: JZ/year.

**Table 1.** Sensitivity experiments used to create the 42 OHCA dataset versions accounting for variations in XBT bias corrections (six corrections plus an uncorrected version) and six mapping methods. All experiments are relative to the monthly mean baseline climatology from Alory et al. (2007) and which does not include XBT data. Mapping methods are from: DOM (Domingues et al. 2008), LEV (Levitus et al. 2012), ISH (Ishii and Kimoto 2009), EN (Ingleby and Huddleston 2007), PMEL (Lyman and Johnson 2008) and WIL (Willis et al. 2004). Periods available are for 1970-2008 or 1993-2008, depending on the mapping method.

| XBT bias corrections | | Mapping Methods | | | | | |
|---|---|---|---|---|---|---|---|
| Acronym/Reference | Correction | DOM | LEV | ISH | EN | PMEL | WIL |
| **No_corr** | No correction | 1970- | 1970- | 1970- | 1970- | 1970- | 1993- |
| **W08** (Wijffels et al. 2008) | Depth | 1970- | 1970- | 1970- | 1970- | 1970- | 1993- |
| **I09** (Ishii and Kimoto, 2009) | Depth | 1970- | 1970- | 1970- | 1970- | 1970- | 1993- |
| **L09** (Levitus et al. 2009) | Temperature | 1970- | 1970- | 1970- | 1970- | 1970- | 1993- |
| **GD11** (Good, 2011) | Depth (bathymetry approach) | 1970- | 1970- | 1970- | 1970- | 1970- | 1993- |
| **GK12** (Gouretski 2012) | Depth + Temperature | 1970- | 1970- | 1970- | 1970- | 1970- | 1993- |
| **CW13** (Cowley et al. 2013) | Depth + Temperature | 1970- | 1970- | 1970- | 1970- | 1970- | 1993- |

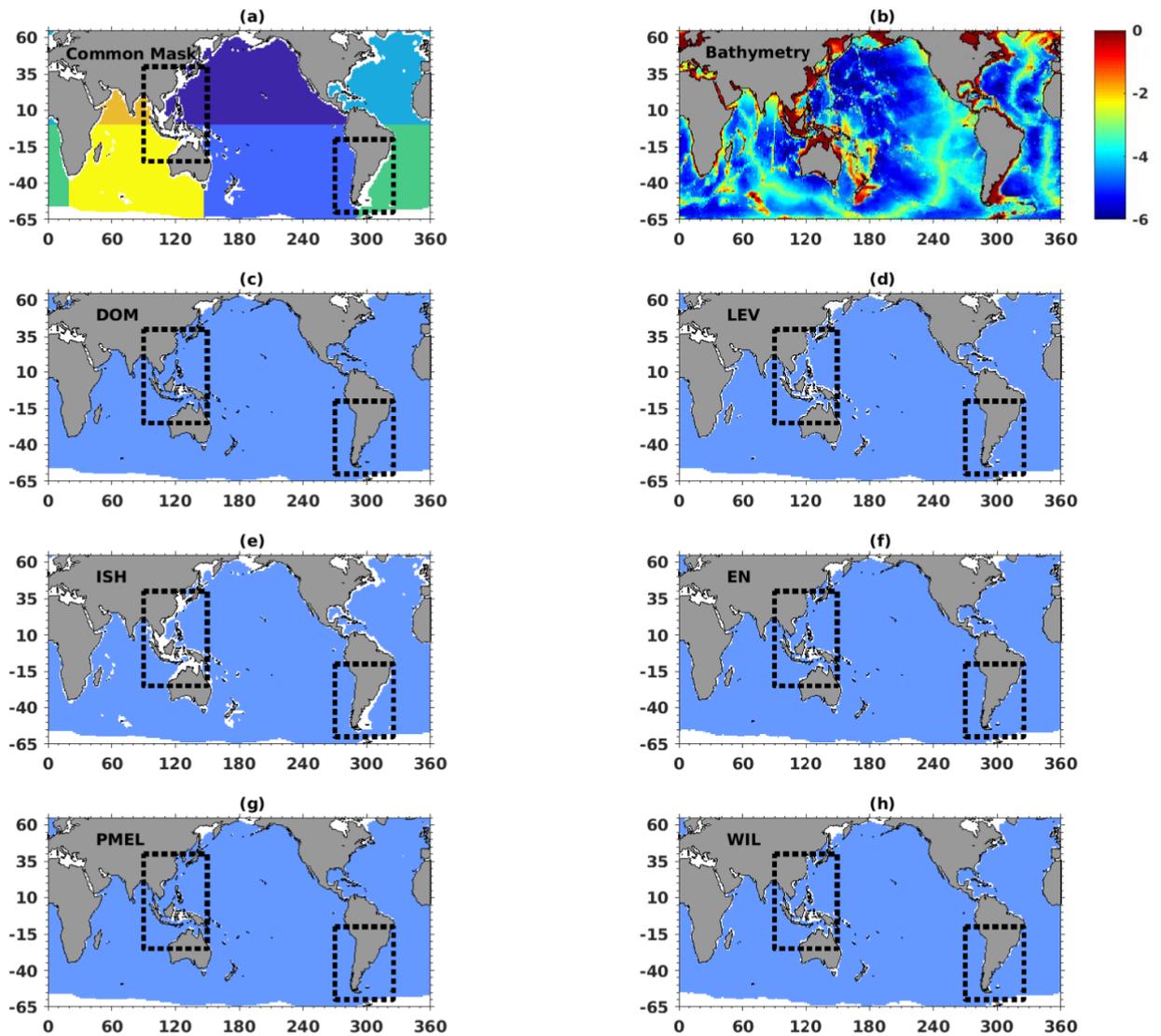

**Fig. 1.** Ocean mask definitions and bathymetry. (a) Common mask for global and basins (color). (b) Bathymetry (km) from ETOPO5 (https://doi.org/10.7289/V5D798BF). (c)–(h) Original masks from DOM, LEV, ISH, EN, PMEL and WIL respectively, where blue denotes where there is data coverage and white where there is no data. Southern Ocean basin is poleward of 35°S (not shown). Boxes (dotted line) illustrate major differences in the original masks among research groups. The total area of common mask (2.86 x $10^{14}$ m$^2$) and individual masks are: DOM (3.22 x$10^{14}$ m$^2$), LEV (3.04 x$10^{14}$ m$^2$), ISH (2.91 x $10^{14}$ m$^2$), EN (3.11 x $10^{14}$ m$^2$), PMEL (3.18 x $10^{14}$ m$^2$) and WIL (3.11 x $10^{14}$ m$^2$).

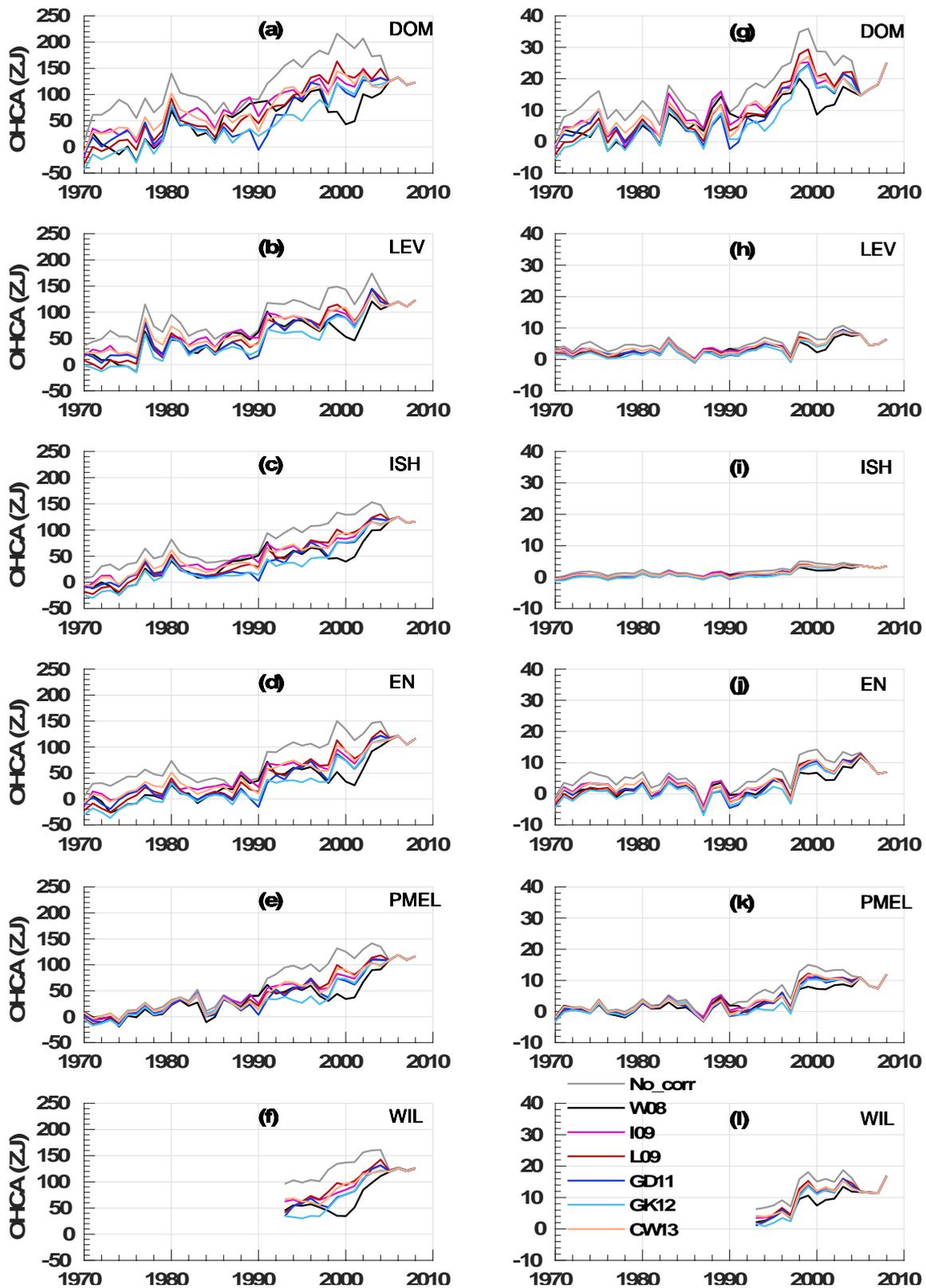

**Fig. 2.** Global OHCA timeseries for six XBT bias corrections as well as uncorrected version (see legend). Left: Timeseries based on the common mask for each mapping method: (a) DOM, (b) LEV, (c) ISH, (d) EN and (e) PMEL for 1970–2008, and (f)

WIL for 1993–2008. Right: Differences based on the original mask minus common mask. Units: ZJ.

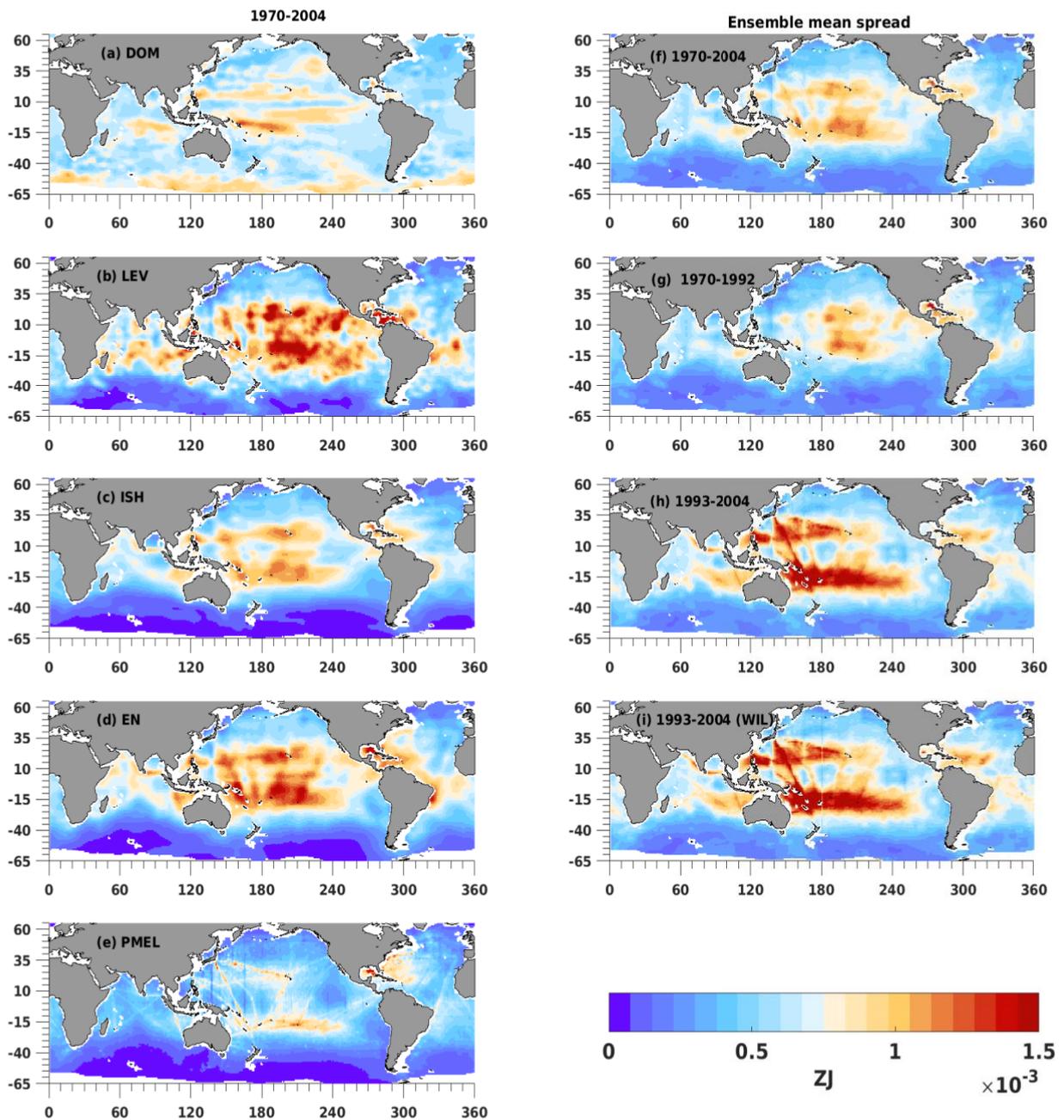

**Fig. 3**. OHCA spread due to XBT bias correction. Left: For each mapping method ((a) DOM, (b) LEV, (c) ISH, (d) EN and (e) PMEL) averaged over 1970–2004. Right: Ensemble mean spread (EnSTD) across mapping methods for different time periods (f) 1970–2004, (g) 1970–1992, (h) 1993–2004 excluding WIL (only available since 1993) and (i) 1993–2004 including WIL. Units: ZJ.

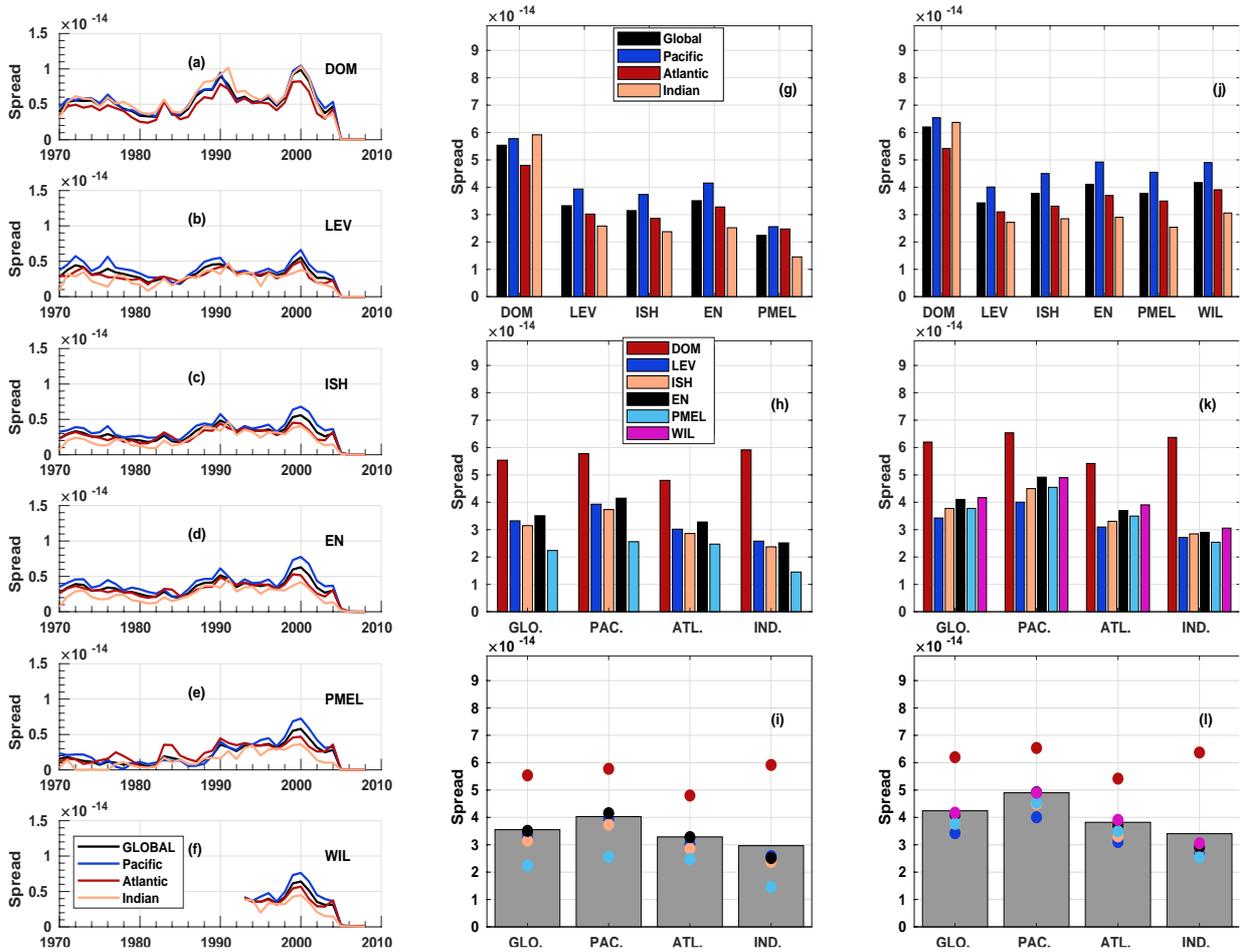

**Fig. 4.** Global and basin OHCA spread due to XBT bias correction per metre square. Left: Annual timeseries from 1970–2008 for each mapping method: (a) DOM, (b) LEV, (c) ISH, (d) EN, (e) PMEL and (f) WIL. Middle: Spread averaged over 1970–2004 organised by (g) mapping method, (h) by basin and (i) ensemble mean spread (EnSTD, gray bars) across mapping methods organized by basin. Right: same as middle but for 1993–2004. Units: ZJ m$^{-2}$.

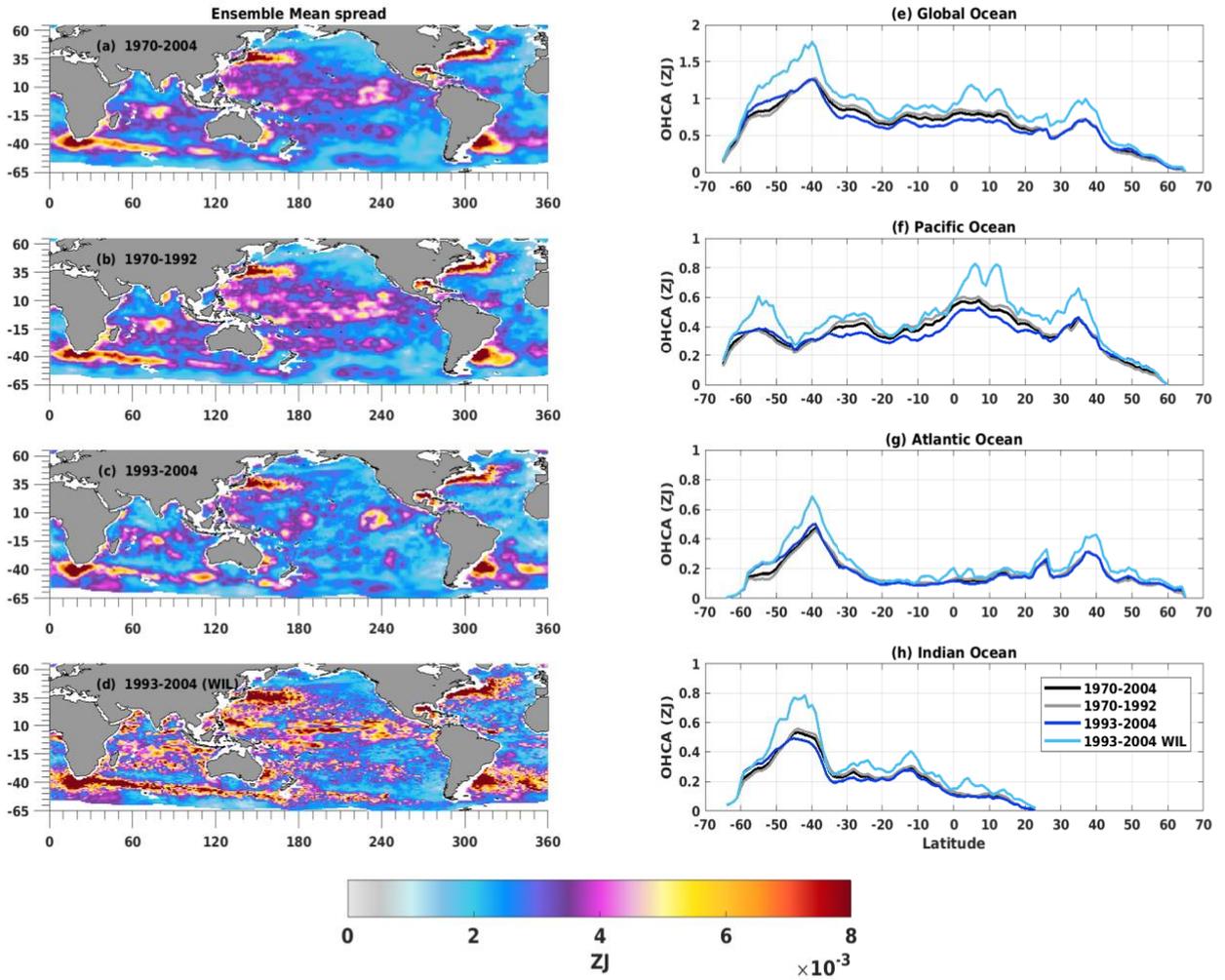

**Fig. 5.** OHCA ensemble mean spread (EnSTD) due to mapping method across XBT bias corrections for different periods. Left: Global patterns for (a) 1970–2004, (b) 1970–1992, (c) 1993–2004 excluding WIL and (d) 1993–2004 including WIL. Right: Zonal integrals for (e) global, (f) Pacific, (g) Atlantic and (h) Indian oceans. Units: ZJ.

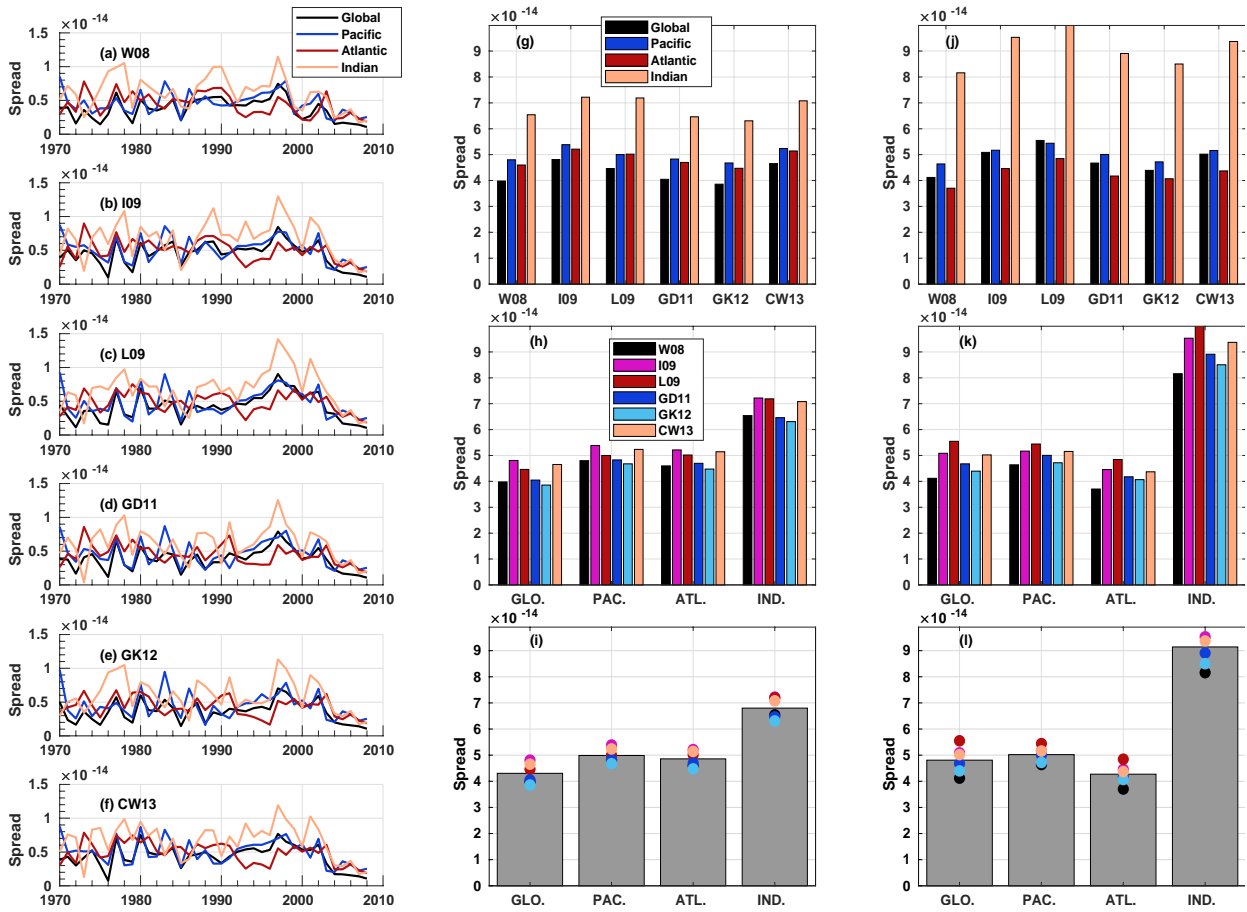

**Fig. 6.** Global and basin OHCA spread due to mapping method per metre square. Left: Annual timeseries from 1970–2008 for each XBT bias correction: W08 (a), I09 (b), L09 (c), GD11 (d), GK12 (e) and CW13 (f). Middle: Spread averaged over 1970–2004 organised by XBT correction (g), by basin (h) and the ensemble mean spread (EnSTD, gray bars) across XBT bias corrections organized by basin (i). Right: Same as middle but for 1993–2004. Units: ZJ m$^{-2}$.

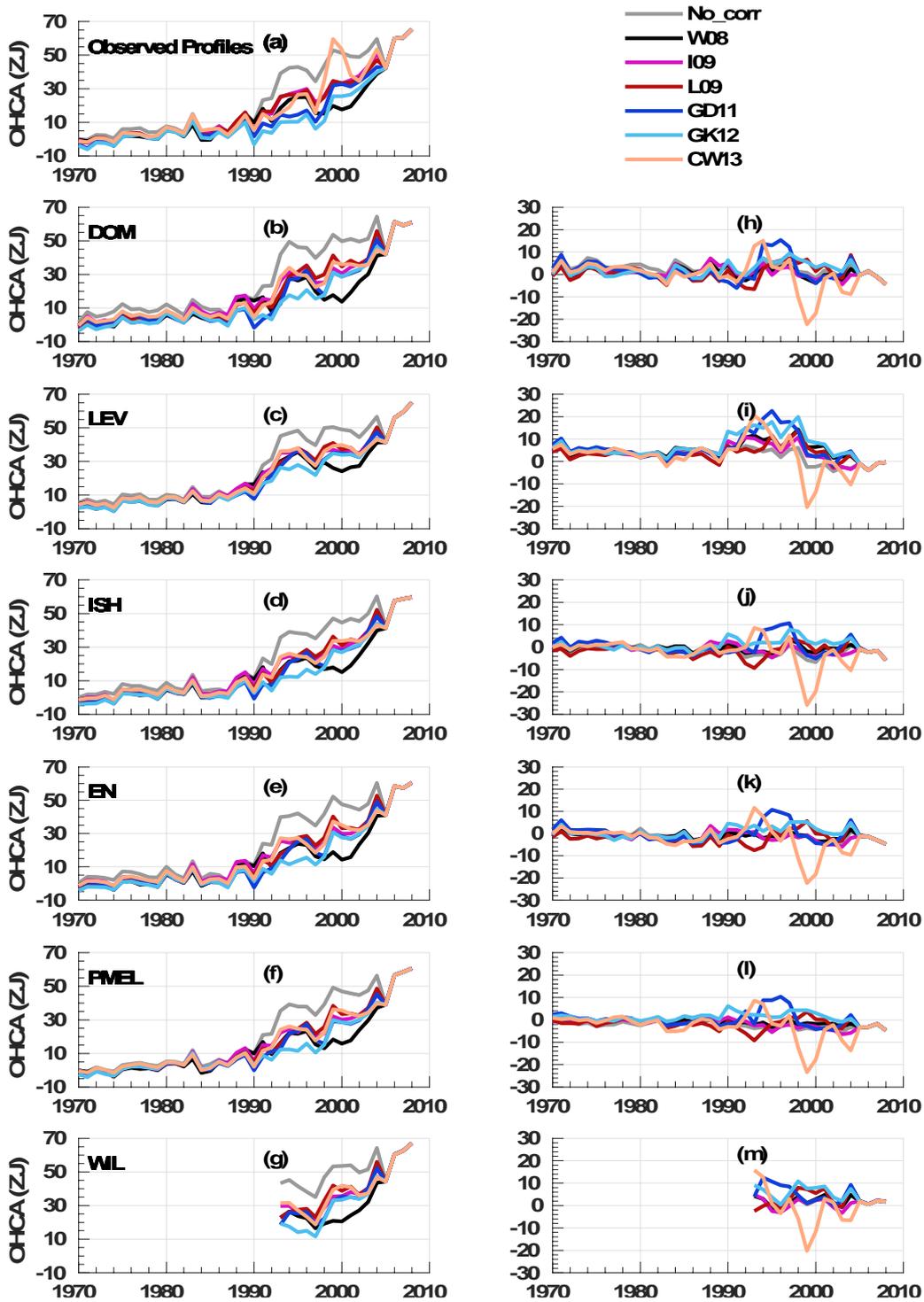

**Fig. 7.** Global OHCA annual timeseries for six XBT bias corrections as well as uncorrected version (see legend) based on subsampled profiles for 0–700 m only. Left: For observed profiles with no mapping applied ("*observed profiles*") and subsampled profiles ("*subsampled*") from each mapping method: (b) DOM, (c) LEV, (d) ISH, (e) EN and (f) PMEL for 1970–2008, and (g) WIL for 1993–2008. Right: Differences based on the original mask minus common mask. Units: ZJ.

**(A) 1970-2004**

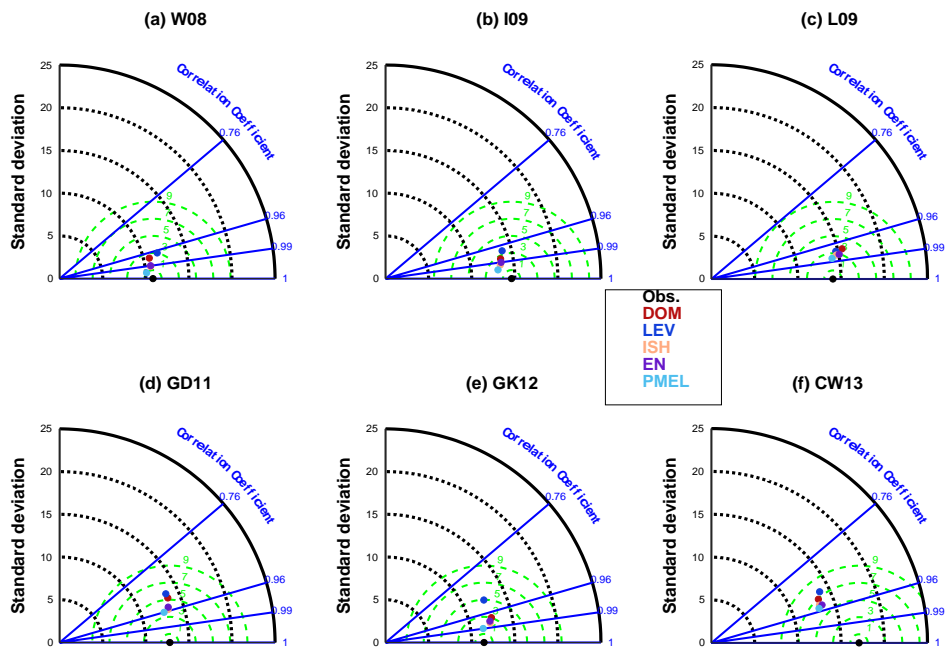

**(B) 1993-2004**

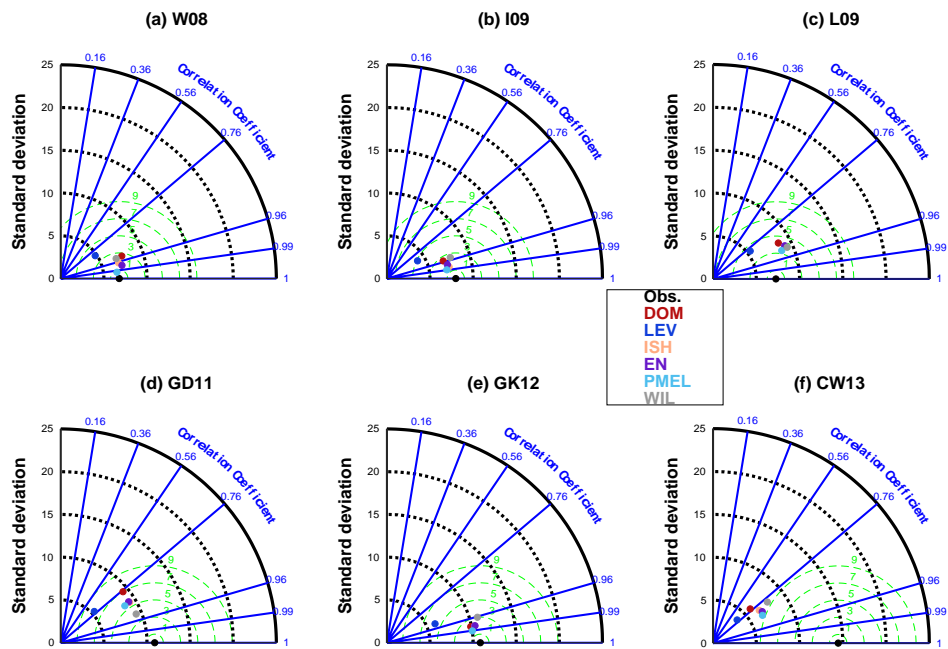

**Fig. 8.** Taylor diagrams for each XBT bias correction comparing "*observed profiles*" for 0–700 m (only) and respective subsampled profiles mapped by six methods (color legend). (A): For 1970–2004. (B): For 1993–2004. XBT bias corrections are: (a) W08, (b) I09, (c) L09, (d) GD11, (e) GK12 and (f) CW13. STD (black axis) and Root Mean Square Error (RMSE, green axis) in ZJ. Correlation coefficients (blue axis) are normalized.

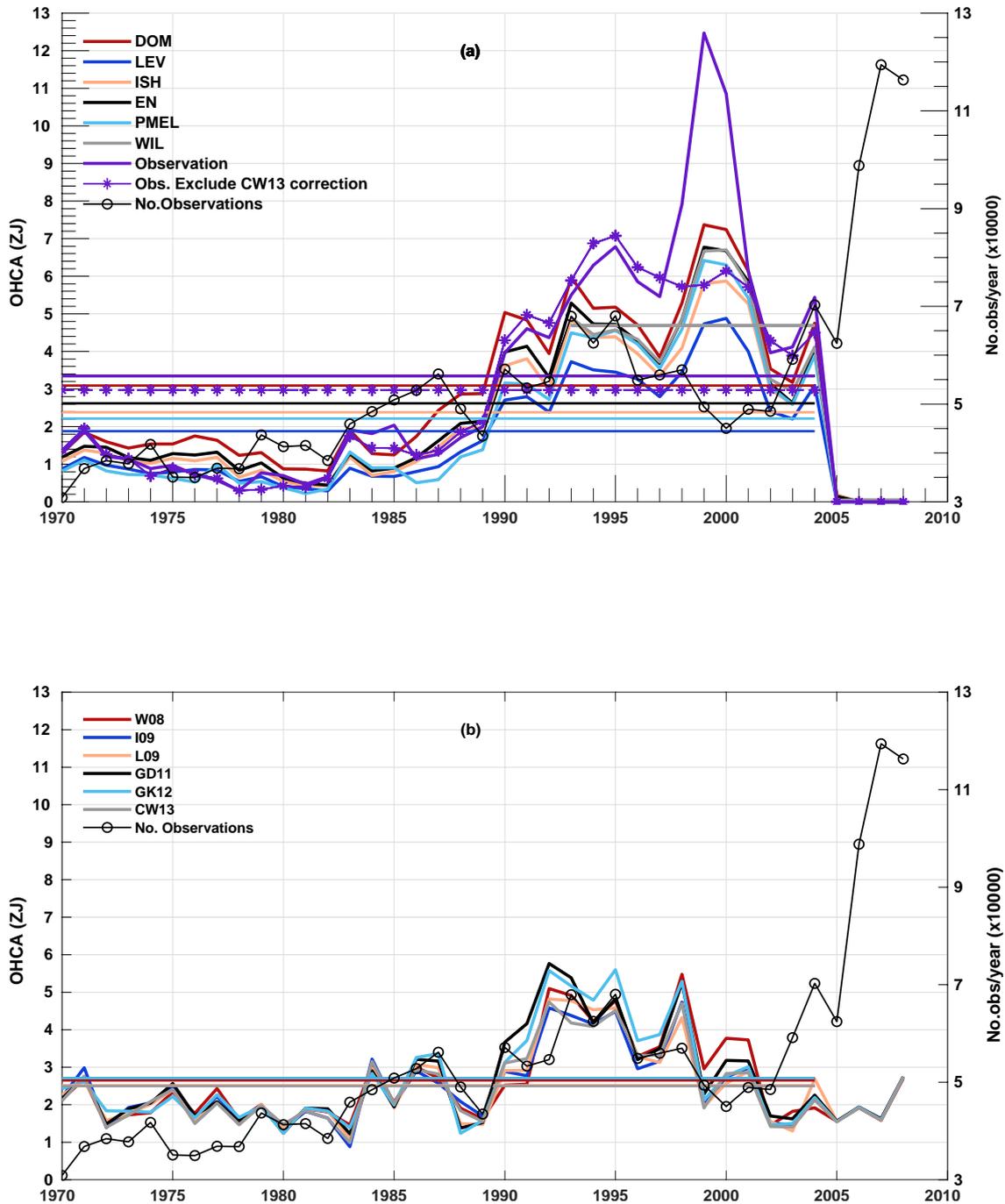

**Fig. 9.** Global OHCA spread timeseries based on subsampled profiles for 0–700 m only (left axis), and profile numbers per year (black line with circle, right axis). (a): Spread due to XBT bias correction across six mapping methods ("*subsampled*") and for observed profiles without mapping ("*observed*"), and with and without the XBT correction from CW13. (b) Spread due to mapping methods across six XBT bias corrections. Horizontal lines represent the time mean for 1970–2004 (or 1993–2004 for WIL). Units: ZJ.

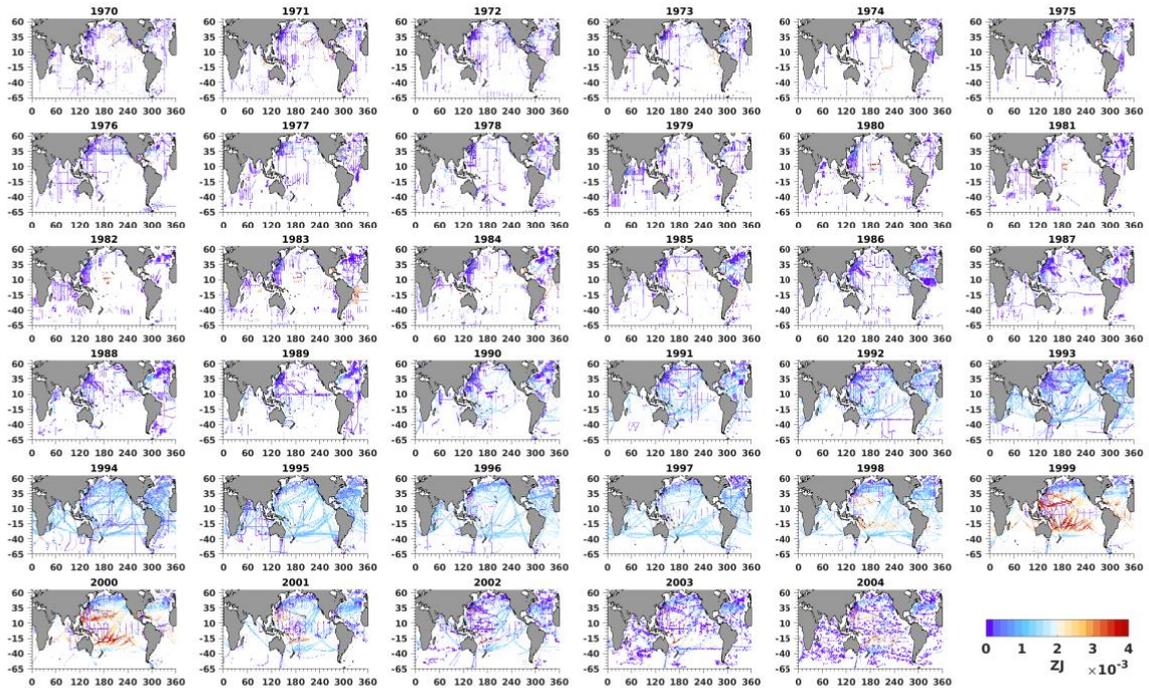

**Fig. 10.** Annual global maps of OHCA spread due to XBT bias correction from observed profiles ("*observed*") for 0–700 m only, from 1970 to 2004. Units: ZJ.

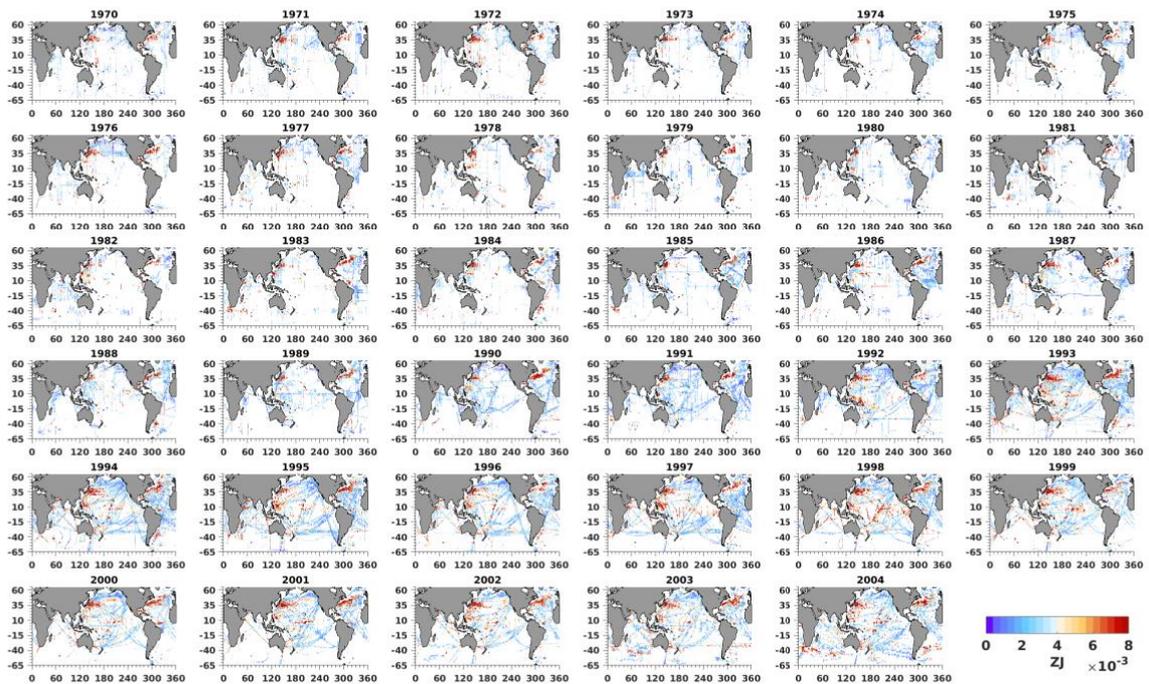

**Fig. 11.** Annual global maps of OHCA spread due to mapping method using the L09 XBT bias correction based on subsampled profiles ("*subsampled*") for 0–700 m only, from 1970 to 2004. Units: ZJ.

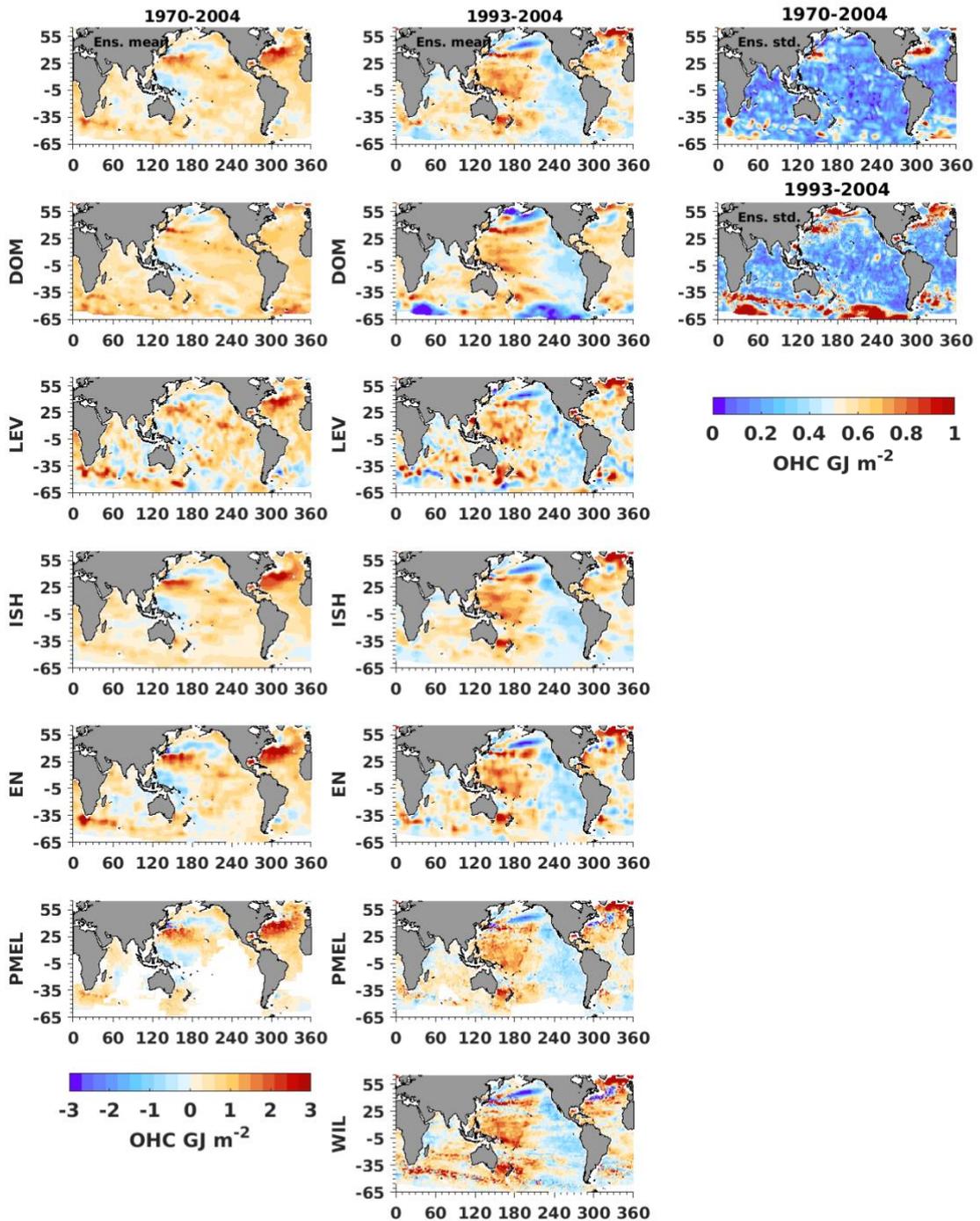

**Fig. 12.** OHCA linear trend maps based on six mapping methods (DOM, LEV, ISH, PMEL and WIL) using the L09 XBT bias correction and their ensemble mean of the mapping methods. Left: for 1970–2004. Middle: for 1993–2004. Right: Ensemble mean spread (EnSTD) for 1970–2004 (top) and for 1993–2004 (bottom). Units: GJ/m$_2$; 1GJ=10$_9$J.

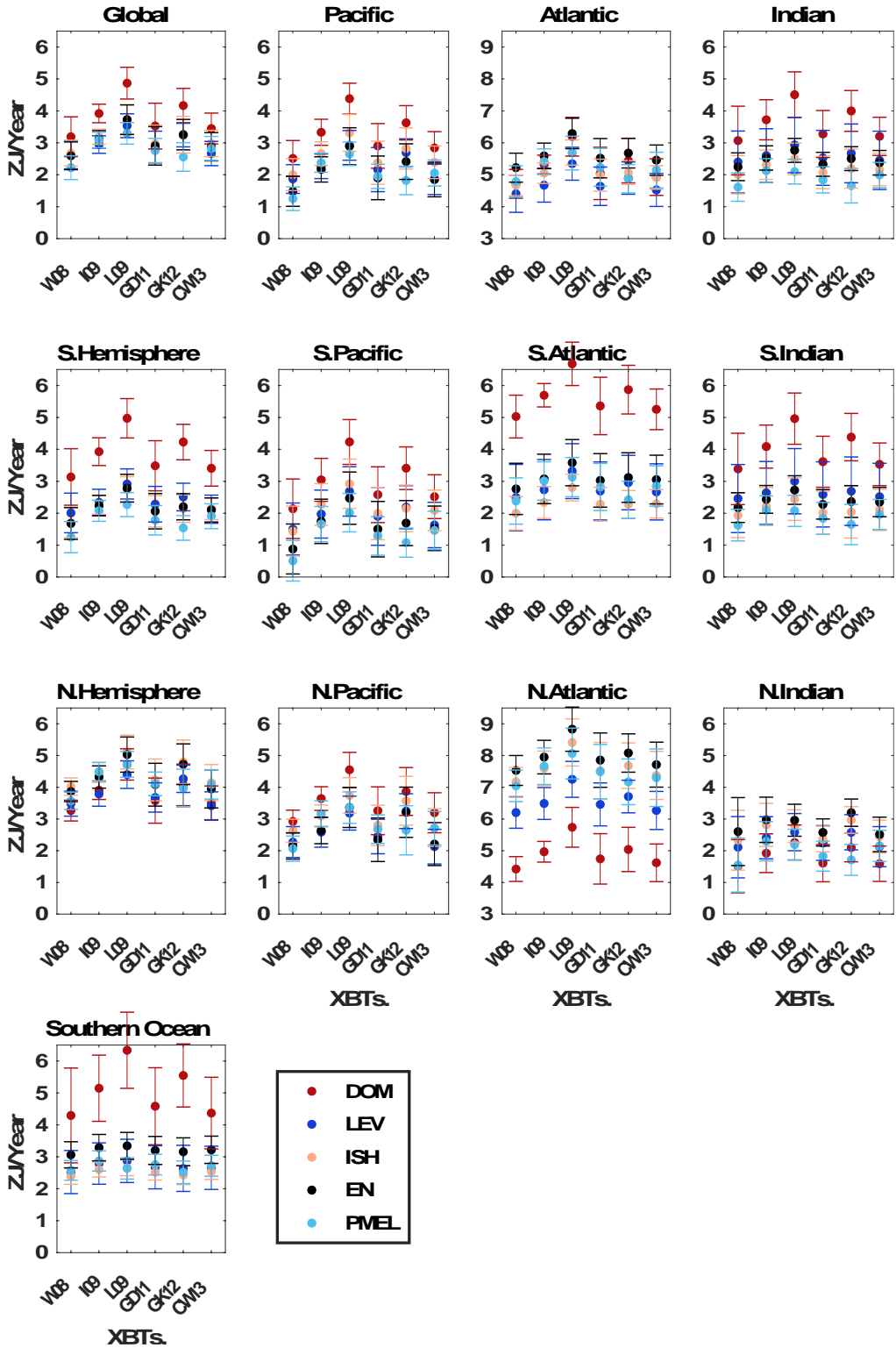

**Fig. 13.** Linear OHCA trends for different basins (panels) and for each XBT bias correction (x-axis) and for each mapping method (color legend) for 1970–2004. Errorbars take into account the reduction in the degrees of freedom due to the temporal correlation in the residuals (**Section 2d**). Unit: JZ/year.

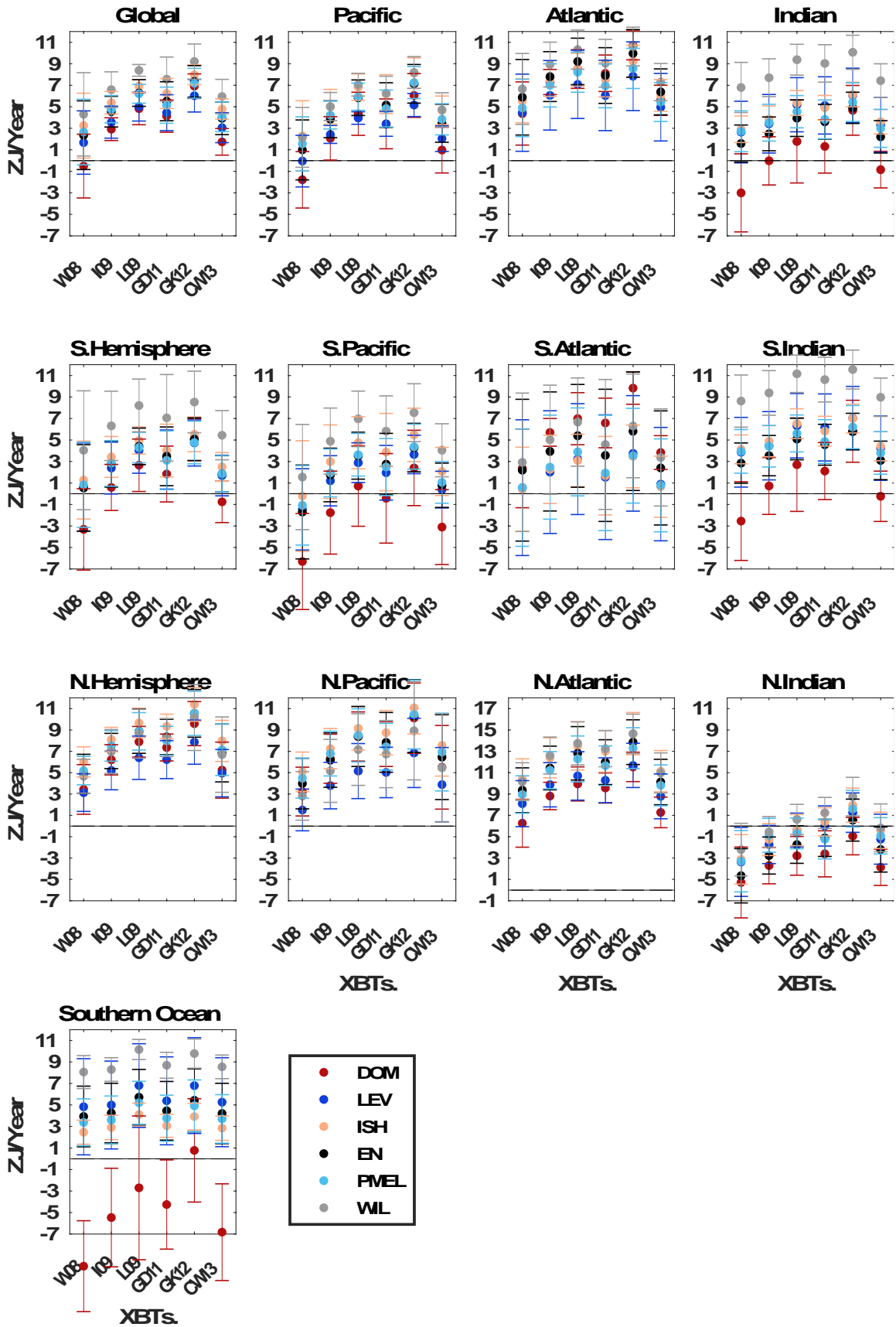

**Fig. 14.** As in Fig. 13 but for 1993–2004, and including WIL (only available since 1993). Unit: JZ/year.